\documentclass[twoside,leqno,twocolumn]{article}
\usepackage{ltexpprt}
\usepackage{enumerate}
\usepackage{t1enc}
\usepackage{pgfplots}
\usepackage{tikz}
\usepackage{amsfonts}

\newcommand{\set}[1]{\left\{ #1\right\}}

\newcommand{\sodass}{\,:\,}
\newcommand{\setGilt}[2]{\left\{ #1\sodass #2\right\}}

\newcommand{\realrange}[2]{\left[#1, #2\right]}

\newcommand{\unitrange}[2]{\realrange{0}{1}}

\newcommand{\llabel}[1]{\label{\labelprefix:#1}}
\newcommand{\labelprefix}{} 

\newcommand{\discussionsize}{\small}

\newcommand{\frage}[1]{}

\newenvironment{code}{\noindent
\begin{tabbing}%
\hspace{2em}\=\hspace{2em}\=\hspace{2em}\=\hspace{2em}\=\hspace{2em}\=%
\hspace{2em}\=\hspace{2em}\=\hspace{2em}\=\hspace{2em}\=\hspace{2em}\=%
\kill}{\end{tabbing}}

\newcommand{\labelcommand}{}
\newcommand{\captiontext}{}
\newsavebox{\codeparam}
\newcounter{lineNumber}
\newenvironment{disscodepos}[3]{%
\renewcommand{\labelcommand}{#2}%
\renewcommand{\captiontext}{#3}%
\sbox{\codeparam}{\parbox{\textwidth}{#3}}%
\begin{figure}[#1]\begin{center}\begin{code}\setcounter{lineNumber}{1}}{%
\end{code}\end{center}\caption{\llabel{\labelcommand}\captiontext}\end{figure}}

{\end{disscodepos}}

\newdimen\endofsize\endofsize=0.5em
\def\endofbeweis{~\quad\hglue\hsize minus\hsize
                 \hbox{\vrule height \endofsize width
\endofsize}\par}

\usepackage{epsfig}
\usepackage{url}
\usepackage{amsmath}
\usepackage{amssymb}
\usepackage{pdflscape}
\usepackage{latexsym}
\usepackage{todonotes}
\usepackage{hyperref}
\usepackage[algo2e,ruled,vlined]{algorithm2e}
\usepackage{algorithmic}
\usepackage{algorithm}
\usepackage{wrapfig}
\usepackage{graphicx}
\usepackage{numprint}
\npdecimalsign{.} 
\usepackage{makeidx}
\usepackage{booktabs}
\newcommand{\Id}[1]{\ensuremath{\text{{\sf #1}}}}
\usepackage{doi}
\newcommand{\ie}{i.e.\ }
\newcommand{\etal}{et~al.}
\newcommand{\eg}{e.g.\ }

\newcommand{\myparagraph}[1]{\vspace{6pt}\par\noindent{\bf#1}}

\usepackage{color}
\usepackage{colortbl}
\usepackage{etoolbox}
\usepackage[left] {lineno}
\definecolor {infocolor} {rgb} {0.6,0.6,0.6}

\newcommand{\I}[0]{{\ensuremath \mathcal{I}}}

\definecolor{addgreen}{rgb}{0.0, 0.7, 0.0}

\newcommand{\longRW}[1]{}

\usepackage[normalem]{ulem}
\definecolor{removered}{rgb}{0.7, 0.0, 0.0}
\newcommand{\remove}[1]{\textcolor{removered}{\sout{#1}}}

\newif\ifFull
\Fullfalse

\newif\ifDoubleBlind
\DoubleBlindtrue

\newif\ifAppendix
\Appendixtrue

\newif\ifFull
\Fullfalse

\let\doendproof\endproof
\renewcommand\endproof{~\hfill$\boxtimes$\doendproof}
\newtheorem{reduction}{Reduction}

\usepackage{csvsimple}

\usepackage[labelformat=simple]{subcaption}

\usepackage{afterpage}

\usepackage[square,numbers,sort&compress]{natbib}

\newcommand{\niceremark}[3]{\textcolor{red}{\textsc{#1 #2: }}\textcolor{blue}{\textsf{#3}}}

\newcommand{\Cminus}{C\raisebox{.08ex}{\hbox{\tt ++}}\raisebox{.25ex}{-}}
\newcommand{\GG}{g\raisebox{.08ex}{\hbox{\tt ++}}}

\newcommand{\BRF}{B~\&~R\textsubscript{\text{full}}}
\newcommand{\BRD}{B~\&~R\textsubscript{\text{dense}}}
\definecolor{lightergray}{rgb}{0.86, 0.86, 0.86}
\definecolor{darkred}{rgb}{0.5, 0.0, 0.0}
\definecolor{darkgreen}{rgb}{0.0, 0.5, 0.0}

\newif\ifRemoveComments
\RemoveCommentsfalse

\ifRemoveComments
\renewcommand{\niceremark}[3]{ }

\renewcommand{\remove}[1]{{}}
\fi

\patchcmd{\thebibliography}{\list}{\fontsize{0.98em}{0.9\baselineskip}\selectfont\list}{}{}

\newcommand{\mytitle}
{Exactly Solving the Maximum Weight Independent Set Problem on Large Real-World Graphs}
\begin{document}
\title{\mytitle\thanks{The research leading to these results has received funding from the European Research Council under the European Union's Seventh Framework Programme (FP/2007-2013) / ERC Grant Agreement no. 340506. This work was partially supported by DFG grants SA 933/10-2.  
}}
\author{Sebastian Lamm\thanks{Institute for Theoretical Informatics, Karlsruhe Institute of Technology, Karlsruhe, Germany. Parts of this research have been done while the author was at the University of Vienna.}, Christian Schulz\thanks{Faculty of Computer Science, University of Vienna, Vienna, Austria.}, Darren Strash\thanks{Department of Computer Science, Hamilton College, Clinton, NY, USA.}, Robert Williger\thanks{Institute for Theoretical Informatics, Karlsruhe Institute of Technology, Karlsruhe, Germany.}, Huashuo Zhang\thanks{Department of Computer Science, Colgate University, Hamilton, NY, USA.}}
  \date{}
\maketitle
\begin{abstract}
One powerful technique to solve NP-hard optimization problems in practice is branch-and-reduce search---which is branch-and-bound that intermixes branching with reductions to decrease the input size. While this technique is known to be very effective in practice for unweighted problems, very little is known for weighted problems, in part due to a lack of known effective reductions. In this work, we develop a full suite of new reductions for the maximum weight independent set problem and provide extensive experiments to show their effectiveness in practice on real-world graphs of up to millions of vertices and edges.

Our experiments indicate that our approach is able to outperform existing state-of-the-art algorithms, solving many instances that were previously infeasible. In particular, we show that branch-and-reduce is able to solve a large number of instances up to two orders of magnitude faster than existing (inexact) local search algorithms---and is able to solve the majority of instances within 15 minutes. For those instances remaining infeasible, we show that combining kernelization with local search produces higher-quality solutions than local search alone.

\end{abstract}
\pagestyle{plain}
\section{Introduction}
The maximum weight independent set problem is an NP-hard problem that has attracted much attention in the combinatorial optimization community, due to its difficulty and its importance in many fields.
Given a graph $G=(V,E,w)$ with weight function $w:V\rightarrow \mathbb{R}^+$, the goal of the maximum weight independent set problem is to compute a set of vertices $\mathcal{I}\subseteq V$ with maximum total weight, such that no vertices in $\mathcal{I}$ are adjacent to one another.
Such a set is called a \emph{maximum weight independent set} (MWIS).
The maximum weight independent set problem has applications spanning many disciplines, including signal transmission, information retrieval, and computer vision~\cite{balas1986finding}.
As a concrete example, weighted independent sets are vital in labeling strategies for maps~\cite{gemsa2014dynamiclabel,barth-2016},
where the objective is to maximize the number of visible non-overlapping labels on a map.
Here, the maximum weight independent set problem is solved in the label conflict graph, where any two overlapping labels are connected by an edge and vertices have a weight proportional to the city's population.
%
%

Similar to their unweighted counterparts, a maximum weight independent set $\mathcal{I}\subseteq V$ in $G$ is a maximum weight clique in $\overline{G}$ (the complement of $G$), and $V \setminus \mathcal{I}$ is a minimum vertex cover of $G$~\cite{xu2016new,cai-dynwvc}.
Since all of these problems are NP-hard~\cite{garey1974}, heuristic algorithms are often used in practice to efficiently compute solutions of high quality on \emph{large} graphs~\cite{pullan-2009,hybrid-ils-2018,li2017efficient,cai-dynwvc}.

Small graphs with hundreds to thousands of vertices may often be solved in practice with traditional branch-and-bound methods~\cite{balas1986finding,babel1994fast,warren2006combinatorial,butenko-trukhanov}. However, even for medium-sized synthetic instances, the maximum weight independent set problem becomes infeasible. Further complicating the matter is the lack of availability of large real-world test instances --- instead, the standard practice is to either systematically or randomly assign weights to vertices in an unweighted graph. Therefore, the performance of exact algorithms on real-world data sets is virtually unknown.

In stark contrast, the unweighted variants can be quickly solved on \emph{large} real-world instances---even with millions of vertices---in practice, by using \emph{kernelization}~\cite{strash-power-2016,chang2017,hespe2018scalable} or the \emph{branch-and-reduce} paradigm~\cite{akiba-tcs-2016}. For those instances that can't be solved exactly, high-quality (and often exact) solutions can be found by combining kernelization with either local search~\cite{dahlum2016,chang2017} or evolutionary algorithms~\cite{redumis-2017}.

These algorithms first remove (or fold) whole subgraphs from the input graph while still maintaining the ability to compute an optimal solution from the resulting smaller instance.
This so-called \emph{kernel} is then solved by an exact or heuristic algorithm.
While these techniques are well understood, and are effective in practice for the unweighted variants of these problems, very little is known about the weighted~problems.


While the unweighted maximum independent set problem has many known reductions, we are only aware of \emph{one} explicitly known reduction for the maximum weight independent set problem: the weighted critical independent set reduction by Butenko and Trukhanov~\cite{butenko-trukhanov}, which has only been tested on small synthetic instances with unit weight (unweighted case). However, it remains to be examined how their weighted reduction performs in practice.
There is only one other reduction-like procedure of which we are aware, 
although it is neither called so directly nor is it explicitly implemented as a reduction. Nogueria~\etal~\cite{hybrid-ils-2018} introduced the notion of a ``$(\omega,1)$-swap'' in their local search algorithm, which swaps a vertex into a solution if its neighbors in the current solution have smaller total weight. This swap is not guaranteed to select a vertex in a true MWIS; however, we show how to transform it into a reduction that does.

\myparagraph{Our Results.}
In this work, we develop a full suite of new reductions for the maximum weight independent set problem and provide extensive experiments to show their effectiveness in practice on real-world graphs of up to millions of vertices and edges.
While existing exact algorithms are only able to solve graphs with hundreds of vertices, our experiments show that our approach is able to exactly solve real-world label conflict graphs with thousands of vertices, and other larger networks with synthetically generated vertex weights---all of which are infeasible for state-of-the-art solvers.
Further, our branch-and-reduce algorithm is able to solve a large number of instances up to two orders of magnitude faster than existing \emph{inexact} local search algorithms---solving the majority of instances within 15 minutes. For those instances remaining infeasible, we show that combining kernelization with local search produces higher-quality solutions than local search alone.

Finally, we develop new \emph{meta} reductions, which are general rules that subsume traditional reductions. We show that weighted variants of popular unweighted reductions can be explained by two general (and intuitive) rules---which use MWIS search as a subroutine. This yields a simple framework covering many~reductions. 

\ifFull
\myparagraph{Organization.}
The rest of the paper is organized as follows.
We begin in Section~\ref{sec:related_work} by highlighting important related work for the maximum weight independent set problem.
This includes exact and heuristic algorithms, as well as hybrid approaches.
Afterwards, we present basic concepts used in our algorithm in Section~\ref{sec:prelim}. This section also contains reductions for that are used for the unweighted maximum independent set in practice in order to make the weighted reductions that we define later more accessible.
We describe the overall structure of our branch-and-reduce framework in Section~\ref{sec:noveltechniques}. The full set of reductions for the weighted maximum independent set problem that are employed by our algorithm are described in Section~\ref{sec:labelofchapterforoutlineofpaper_reductions}.
An extensive experimental evaluation of our method is presented in Section~\ref{sec:experiments}.
Finally, we present conclusions in Section~\ref{sec:conclusion}.
\fi{}

\section{Related Work}
\label{sec:related_work}
We now present important related work on finding high-quality weighted independent sets.
This includes exact branch-and-bound algorithms, reduction based approaches, as well as inexact heuristics, \eg local search algorithms.
We then highlight some recent approaches that combine both exact and inexact~algorithms.

\subsection{Exact Algorithms.}
Much research has been devoted to improve exact branch-and-bound algorithms for the MWIS and its complementary problems.
These improvements include different pruning methods and sophisticated branching schemes~\cite{ostergaard2002fast,balas1986finding,babel1994fast,warren2006combinatorial}.

Warren and Hicks~\cite{warren2006combinatorial} proposed three combinatorial branch-and-bound algorithms that are able to quickly solve DIMACS and weighted random graphs.
These algorithms use weighted clique covers  to generate upper bounds that reduce the search space via pruning.
Furthermore, they all use a branching scheme proposed by Balas and Yu~\cite{balas1986finding}.
In particular, their first algorithm is an extension and improvement of a method by Babel~\cite{babel1994fast}.
Their second one uses a modified version of the algorithm by Balas and Yu that uses clique covers that borrow structural features from the ones by Babel~\cite{babel1994fast}.
Finally, their third approach is a hybrid of both previous algorithms.
Overall, their algorithms are able to quickly solve instances with hundreds of vertices.

An important technique to reduce the base of the exponent for exact branch-and-bound algorithms are so-called \emph{reduction rules}.
Reduction rules are able to reduce the input graph to an irreducible \emph{kernel} by removing well-defined subgraphs.
This is done by selecting certain vertices that are provably part of some maximum(-weight) independent set, thus maintaining optimality.
We can then extend a solution on the kernel to a solution on the input graph by undoing the previously applied reductions.
There exist several well-known reduction rules for the unweighted vertex cover problem (and in turn for the unweighted MIS problem)~\cite{akiba-tcs-2016}.
However, there are only a few reductions known for the MWIS problem.

One of these was proposed by Butenko and Trukhanov~\cite{butenko-trukhanov}.
In particular, they show that every critical weighted independent set is part of a maximum weight independent set.
A critical weighted set is a subset of vertices such that the difference between its weight and the weight of its neighboring vertices is maximal for all such sets.
They can be found in polynomial time via minimum cuts.
Their neighborhood is recursively removed from the graph until the critical set is~empty.

As noted by Larson~\cite{larson-2007}, it is possible that in the unweighted case the initial critical set found by Butenko and Trukhanov might be empty.
To prevent this case, Larson~\cite{larson-2007} proposed an algorithm that finds a \emph{maximum} (unweighted) critical independent set.
\longRW{His algorithm accumulates vertices that are in some critical set and removes their neighborhood.
Additionally, he provides a method to quickly check if a given vertex is part of some critical set.}
Later Iwata~\cite{iwata-2014} has shown how to remove a large collection of vertices from a maximum matching all at once; however, it is not known if these reductions are equivalent.

For the maximum weight clique problem, Cai and Lin~\cite{cai2016fast} give an exact branch-and-bound algorithm that interleaves between clique construction and reductions.
\longRW{In particular, their algorithm picks different starting vertices to form a clique and then maintains a candidate set to iteratively extend this clique.
In each iteration, the vertex to be added is selected using a benefit estimation function and a dynamic best from multiple selection heuristic~\cite{cai2015balance}.
Once the candidate set is empty, the new solution is compared to the best solution found so far.
If an improvement is found, their algorithm then tries to apply reductions and reduce the graph size.
To be more specific, they use two reduction rules that are able to remove a vertex $v$ by computing upper bounds related to the weight of different neighborhoods of $v$.} 
We briefly note that their algorithm and reductions are targeted at sparse graphs, and therefore their reductions would likely work well for the MWIS problem on \emph{dense} graphs---but not on the sparse graphs we consider here.

\subsection{Heuristic Algorithms.}
Heuristic algorithms such as local search work by maintaining a single solution that is gradually improved through a series of vertex deletions, insertion and swaps.
Additionally, \emph{plateau search} allows these algorithms to explore the search space by performing node swaps that do not change the value of the objective function.
We now cover state-of-the-art heuristics for both the unweighted and weighted maximum independent set problem.

For the unweighted case, the iterated local search algorithm by Andrade~\etal~\cite{andrade-2012} (ARW) is one of the most successful approaches in practice.
Their algorithm is based on finding improvements using so-called $(1,2)$-swaps that can be found in linear time.
Such a swap removes a single vertex from the current solution and inserts two new vertices instead.
Their algorithm is able to find (near-)optimal solutions for small to medium-sized instances in milliseconds, but
struggles on massive instances with millions of vertices and edges~\cite{dahlum2016}.
\longRW{Their algorithm was improved significantly by applying (2,1)- and ($k$,1)-swaps as perturbation steps~\cite{jin-hao-swap-2015}, omitting high-degree vertices~\cite{dahlum2016} and using reduction rules~\cite{dahlum2016,redumis-2017}.}

\longRW{As most other local search algorithm PLS interleaves sequences of iterative improvements and plateau search.
Improvements are found using different sub-algorithms that select vertices either at random, by using their degree or by using a penalty strategy that is dynamically adjusted during the main algorithm.
Additionally, these sub-algorithms differ in their perturbation mechanisms.
PLS achieved state-of-the-art quality on various benchmark sets including the popular DIMACS benchmark.}

Several local search algorithms have been proposed for the maximum weight independent set problem.
Most of these approaches interleave a sequence of iterative improvements and plateau search.
Further strategies developed for these algorithms include the usage of sub-algorithms for vertex selection~\cite{pullan-2006,pullan-2009}, tabu mechanisms using randomized restarts~\cite{wu2012multi}, and adaptive perturbation strategies~\cite{benlic2013breakout}.
Local search approaches are often able to obtain high-quality solutions on medium to large instances that are not solvable using exact algorithms.
Next, we cover some of the most recent state-of-the-art local search algorithms in greater detail.

The hybrid iterated local search (HILS) by Nogueria~\etal~\cite{hybrid-ils-2018} extends ARW to the weighted case. 
It uses two efficient neighborhood structures: $(\omega, 1)$-swaps and weighted $(1,2)$-swaps.
Both of these structures are explored using a variable neighborhood descent procedure.
\longRW{Additionally, they introduce a dynamic perturbation mechanism that regulates intensification and diversification.}
Their algorithm outperforms state-of-the-art algorithms on well-known benchmarks and is able to find known optimal solution in milliseconds.

Recently, Cai~\etal~\cite{cai-dynwvc} proposed a heuristic algorithm for the weighted vertex cover problem that was able to derive high-quality solution for a variety of large real-world instances.
Their algorithm is based on a local search algorithm by Li~\etal~\cite{li2017efficient} and uses iterative removal and maximization of a valid vertex cover.
\longRW{Vertices are removed and inserted using two scoring function: a \emph{gain} function and a \emph{loss} function.
In particular, the baseline algorithm removes two vertices in each iteration.
The first vertex is chosen using the minimal loss value, the second one is selected by the best from multiple selection heuristic~\cite{cai2015balance} (w.r.t. the minimal loss value).
They then introduce two dynamic approaches for selecting between different scoring functions for the gain and loss values of a vertex.
To be more precise, their first approach selects between two different scoring functions by counting the number of non-improving steps.
Their second approach, dynamically adjust the number of vertices which are removed in a single iteration based on their total degree.}

\subsection{Hybrid Algorithms.}
In order to overcome the shortcomings of both exact and inexact methods, new approaches that combine reduction rules with heuristic local search algorithms were proposed recently~\cite{dahlum2016,redumis-2017}.
A very successful approach using this paradigm is the reducing-peeling framework proposed by Chang~\etal~\cite{chang2017} which is based on the techniques proposed by Lamm~\etal~\cite{redumis-2017}.
Their algorithm works by computing a kernel using practically efficient reduction rules in linear and near-linear time.
Additionally, they provide an extension of their reduction rules that is able to compute good initial solutions for the kernel.
In particular, they greedily select vertices that are unlikely to be in a large independent set, thereby opening up the reduction space again.
Thus, they are able to significantly improve the performance of the ARW local search algorithm that is applied on the kernelized graph.
To speed-up kernelization, Hespe~\etal~\cite{hespe2018scalable} proposed a shared-memory  algorithm using partitioning and parallel bipartite matching. 
\longRW{Additionally, they propose two pruning techniques to provide additional performance gains over the algorithm by Akiba and Iwata~\cite{akiba-tcs-2016}.
These techniques are called dependency checking and reduction tracking.
Dependency checking allows them to prune reductions when they will provably not succeed, therefore significantly reducing the number of failed reductions.
Reduction tracking enables them to stop local reductions when they are not effectively reducing the global graph sizes.
However, using reduction tracking does not produce a full kernel.}

\section{Preliminaries}
\label{sec:prelim}
Let  $G=(V=\{0,\ldots, n-1\},E, w)$ be an undirected graph with $n = |V|$ nodes and $m = |E|$ edges.
$w:V \rightarrow \mathbb{R}^+$ is the real-valued vertex weighting function such that $w(v) \in \mathbb{R}^+$ for all $v \in V$.
Furthermore, for a non-empty set $S \subseteq V$ we use $w(S) = \sum_{v\in S} w(v)$ and $|S|$ to denote the \emph{weight} and \emph{size} of $S$.
The set $N(v) = \setGilt{u}{\set{v,u}\in E}$ denotes the neighbors of $v$.
We further define the neighborhood of a set of nodes $U \subseteq V$ to be $N(U) = \cup_{v\in U} N(v)\setminus U$,
$N[v] = N(v) \cup \{v\}$, and $N[U] = N(U) \cup U$.
A graph $H=(V_H, E_H)$ is said to be a \emph{subgraph} of $G=(V, E)$ if $V_H \subseteq V$ and $E_H \subseteq E$.
We call $H$ an \emph{induced} subgraph when $E_H = \setGilt{\{u,v\} \in E}{u,v\in V_H}$.
For a set of nodes $U\subseteq V$, $G[U]$ denotes the subgraph induced by $U$.
The \emph{complement} of a graph is defined as $\overline{G} = (V,\overline{E})$, where $\overline{E}$ is the set of edges not present in $G$.
An \emph{independent set} is a set $\mathcal{I} \subseteq V$, such that all nodes in $\mathcal{I}$ are pairwise non-adjacent.
An independent set is \emph{maximal} if it is not a subset of any larger independent set.
Furthermore, an independent set $\I$ has \emph{maximum weight} if there is no heavier independent set, \ie~there exists no independent set $I'$ such that $w(I) < w(I')$.

The weight of a maximum independent set of $G$ is denoted by $\alpha_w(G)$.
The \emph{maximum weight independent set problem} (MWIS) is that of finding the independent set of largest weight among all possible independent sets.
A \emph{vertex cover} is a subset of nodes $C \subseteq V$, such that every edge $e \in E$ is incident to at least one node in $C$.
The \emph{minimum-weight vertex cover problem} asks for the vertex cover with the minimum total weight.
Note that the vertex cover problem is complementary to the independent set problem, since the complement of a vertex cover $V \setminus C$ is an independent set. Thus, if $C$ is a minimum vertex cover, then $V \setminus C$ is a maximum independent set.
A \emph{clique} is a subset of the nodes $Q \subseteq V$ such that all nodes in $Q$ are pairwise adjacent.
An independent set is a clique in the complement graph.

\subsection{Unweighted Reductions.}
\label{subsec:reductions}
In this section, we describe reduction rules for the \emph{unweighted} maximum independent set problem.
These reductions perform exceptionally well in practice and form the basis of our weighted reductions described~in~Section~\ref{sec:new-reductions}.

\myparagraph{Vertex Folding~\cite{chen1999}.}
Vertex folding was first introduced by Chen et al.~\cite{chen1999} to reduce the theoretical running time of exact branch-and-bound algorithms for the maximum independent set problem.
This reduction is applied whenever there is a vertex $v$ with degree two and non-adjacent neighbors $u$ and $w$.
Chen et al.~\cite{chen1999} then showed that either $v$ or both $u$ and $w$ are in some maximum independent set.
Thus, we can contract $u$, $v$, and $w$ into a single vertex $v'$ (called a \emph{fold}), forming a new graph $G'$. Then $\alpha(G) = \alpha(G') + 1$
and after finding a MIS $\I'$ of $G'$, 
if $v'\notin \I'$ then $\I = \I'\cup\{v\}$ is an MIS of $G$, otherwise $\I = (\I'\setminus\{v'\})\cup \{u,w\}$ is.

\myparagraph{Isolated Vertex Removal~\cite{butenko-correcting-codes-2009}.}
An \emph{isolated} vertex, also called a \emph{simplicial} vertex, is a vertex $v$ whose neighborhood forms a clique.
That is, there is a clique $C$ such that $V(C) \cap N[v] = N[v]$; this clique is called an \emph{isolated clique}.
Since $v$ has no neighbors outside of the clique, by a cut-and-paste argument, it must be in \emph{some} maximum independent set.
Therefore, we can add $v$ to the maximum independent set we are computing, and remove $v$ and $C$ from the graph.
Isolated vertex removal was shown by Butenko~\etal~\cite{butenko-correcting-codes-2009} to be highly effective in finding exact maximum independent sets on graphs derived from error-correcting codes.
In order to work efficiently in practice, this reduction is typically limited to cliques with size at most 2 or 3~\cite{chang2017,dahlum2016}. 

Although Chang~\etal~\cite{chang2017} showed that the domination reduction (described below) captures the isolated vertex removal reduction, that reduction must be applied several times: once per neighbor in the clique.

\myparagraph{Twin.}
Two non-adjacent vertices $u$ and $v$ are called \emph{twins} if $N(u) = N(v)$. 
Note that either both $u$ and $v$ are in some MIS, or some subset of $N(u)$ is in some MIS.
If $|N(u)| = |N(v)| = 3$, then either $u$ and $v$ are together or \emph{at least} two vertices of $N(u)$ must be in an MIS. The following case of the reduction is relevant to our result: If $N(u)$ is independent, then we can fold $u$, $v$, and $N(v)$ into a single vertex $v'$ and $\alpha(G) = \alpha(G') + 2$.%

\myparagraph{Domination~\cite{fomin-2009}.}
Given two vertices $u$ and $v$, $u$ is said to \emph{dominate} $v$ if and only if $N[u] \supseteq N[v]$. In this case there is an MIS in $G$ that excludes $u$ and therefore, $u$ can be removed from the graph.

\myparagraph{Critical Independent Set.}
A subset $U_c \subseteq V$ is called a \emph{critical set} if $|U_c| - |N(U_c)| = \max\{|U| - |N(U)| : U \subseteq V\}$. 
Likewise, an independent set $I_c \subseteq V$ is called a \emph{critical independent set} if $|I_c| - |N(I_c)| = \max\{|I| - |N(I)| : I \text{ is an independent set of } G\}$.
Butenko and Trukhanov~\cite{butenko-trukhanov} show that any critical independent set is a subset of a maximum independent set.
They then continue to develop a reduction that uses critical independent sets which can be computed in polynomial time.
In particular, they start by finding a critical set in $G$ by using a reduction to the maximum matching problem in a bipartite graph~\cite{ageev1994finding} .
In turn, this problem can then be solved in $\mathcal{O}(|V|\sqrt{|E|})$ time using the Hopcroft-Karp algorithm.
They then obtain a critical independent set by setting $I_c = U_c \setminus N(U_c)$.
Finally, they can remove $I_c$ and $N(I_c)$ from $G$.

\myparagraph{Linear Programming (LP) Relaxation.}
The LP-based reduction rule 
by Nemhauser and Trotter~\cite{nemhauser-1975}, is based on an LP relaxation of the vertex cover problem:
\begin{align*}
  \text{minimize } & \sum_{v \in V} x_v & \\
  \text{s.t. } & x_u + x_v \geq 1 & \text{ for } (u,v) \in E, \\
               & x_v \geq 0 & \text{ for } v \in V.
\end{align*}

Nemhauser and Trotter~\cite{nemhauser-1975} showed that there exists an optimal half-integral solution for this problem.
Additionally, they prove that if a variable $x_v$ takes an integer value in an optimal solution, then there exists an optimal integer solution where $x_v$ has the same value.
Just as in the critical set reduction, they use a reduction to the maximum bipartite matching problem to compute a half-integral solution.
To develop a reduction rule for the vertex cover problem, they afterwards fix the integral part of their solution and output the remaining graph.
Their approach was successively improved by Iwata~\etal~\cite{iwata-2014} 
and was shown to be effective in practice by Akiba and Iwata~\cite{akiba-tcs-2016}.

\subsection{Critical Weighted Independent Set Reduction.}
We now briefly describe the critical weighted independent set reduction, which is the \emph{only} reduction that has appeared in the literature for the \emph{weighted} maximum independent set problem.
Similar to the unweighted case, a subset $U_c \subseteq V$ is called a \emph{critical weighted set} if $w(U_c) - w(N(U_c)) = \max\{w(U) - w(N(U)) : U \subseteq V\}$.
A weighted independent set $I_c \subseteq V$ is called a \emph{critical weighted independent set} (CWIS) if $w(I_c) - w(N(I_c)) = \max\{w(I) - w(N(I)) : I \text{ is an independent set of } G\}$.
Butenko and Trukhanov~\cite{butenko-trukhanov} show that any CWIS is a subset of a maximum weight independent set.
Additionally, they propose a weighted critical set reduction which works similar to its unweighted counterpart.
However, instead of computing a maximum matching in a bipartite graph, a critical weighted set is obtained by solving the selection problem~\cite{ageev1994finding}.
The problem is equivalent to finding a minimum cut in a bipartite graph. 
For a proof of correctness, see the paper by Butenko and Trukhanov~\cite{butenko-trukhanov}.

\begin{reduction}[CWIS Reduction]
Let $U\subseteq V$ be a critical weighted independent set of $G$. Then $U$ is in some MWIS of $G$. We set $G' = G[V\setminus N[U]]$ and $\alpha_w(G) = \alpha_w(G') + w(U)$.
\end{reduction}

\section{Efficient Branch-and-Reduce}
\label{sec:noveltechniques}
We now describe our branch-and-reduce framework in full detail. This includes the pruning and branching techniques that we use, as well as other algorithm details.
An  overview of our algorithm can be found in Algorithm~\ref{branchreducelabel}.
To keep the description simple, the pseudocode describes the algorithm such that it outputs the weight of a maximum weight independent set in the graph. However, our algorithm is implemented to actually output the maximum weight independent set. Throughout the algorithm we maintain the current solution weight as well as the best solution weight. 
Our algorithm applies a set of reduction rules before branching on a node. We describe these reductions in the following section. Initially, we run a local search algorithm on the reduced graph to compute a lower bound on the solution weight, which later helps pruning the search space. 
We then prune the search by excluding unnecessary parts of the branch-and-bound tree to be explored. If the graph is not connected, we separately solve each connected component.
If the graph is connected, we branch  into two cases by applying a branching rule.
If our algorithm does not finish with a certain time limit, we use the currently best solution and improve it using a greedy algorithm. 
More precisely, our algorithm sorts the vertices in decreasing order of their weight and adds vertices in that order if feasible.
We give a detailed description of the subroutines of our~algorithm~below.

\subsection{Incremental Reductions.}
Our algorithm starts by running all reductions that are described in the following section. 
Following the lead of previous works~\cite{strash-power-2016,chang2017,hespe2018scalable}, we apply our reductions \emph{incrementally}. 
For each reduction rule, we check if it is applicable to any vertex of the graph. After the checks for the current reduction are completed, we continue with the next reduction if the current reduction has not changed the graph. If the graph was changed, we go back to the first reduction rule and repeat.
Most of the reductions we introduce in the following section are \emph{local}: if a vertex changes, then we do not need to check the entire graph to apply the reduction again, we only need to consider the vertices whose neighborhood has changed since the reduction was last applied.
The critical weighted independent set reduction defined above is the only \emph{global} reduction that we use; it always considers all vertices in~the~graph.

For each of the local reductions there is a queue of changed vertices associated. Every time a node or its neighborhood is changed it is added to the queues of all reductions. When a reduction is applied only the vertices in its associated queue have to be checked for applicability. After the checks are finished for a particular reduction its queue is cleared. Initially, the queues of all reductions are filled with every vertex~in~the~graph.

\begin{algorithm}[t]
\SetAlgoLined
\begin{algorithmic}
	\STATE   \textbf{input} graph $G=(V,E)$, current solution weight $c$ (initially zero), best solution weight $\mathcal{W}$ (initially zero)
        \vspace*{-.45cm}
	\STATE   \textbf{procedure} Solve($G$, $c$, $\mathcal{W}$)
	\STATE   \quad $(G,c) \leftarrow$ Reduce$(G, c)$
	\STATE   \quad \textbf{if} $\mathcal{W} = 0$ \textbf{then} $\mathcal{W}  \leftarrow$ $c+\mathrm{ILS}(G)$ 
	\STATE   \quad \textbf{if} $c$ + UpperBound($G$) $\leq \mathcal{W}$ \textbf{then} \textbf{return} $\mathcal{W}$
	\STATE   \quad \textbf{if} $G$ is empty \textbf{then} \textbf{return} $\max\{\mathcal{W}, c\}$
	\STATE   \quad \textbf{if} $G$ is not connected \textbf{then}
	\STATE   \quad \quad \textbf{for all} $G_i \in $ Components($G$) \textbf{do}
	\STATE   \quad \quad \quad $c \leftarrow c + \text{Solve}$($G_i$, 0, 0)
	
	\STATE   \quad \quad \textbf{return} $\max(\mathcal{W},c)$
	\STATE   \quad $(G_1, c_1), (G_2, c_2) \leftarrow $ Branch$(G, c)$
	\STATE   \quad  \COMMENT{Run 1st case, update currently best solution}
	\STATE   \quad $\mathcal{W} \leftarrow $ Solve$(G_1, c_1, \mathcal{W})$ 
	\STATE   \quad \COMMENT{Use updated $\mathcal{W}$ to shrink the search space}
	\STATE   \quad $\mathcal{W}\leftarrow $ Solve$(G_2, c_2, \mathcal{W})$
	\STATE   \textbf{return} $\mathcal{W}$
\end{algorithmic}

\caption{Branch-and-Reduce Algorithm for MWIS}
\label{branchreducelabel}
\end{algorithm}

\subsection{Pruning.}
Exact branch-and-bound algorithms for the MWIS problem often use weighted clique covers to compute an upper bound for the optimal solution~\cite{warren2006combinatorial}.
A weighted clique cover of $G$ is a collection of (possibly overlapping) cliques $C_1, \ldots, C_k \subseteq V$, with associated weights $W_1, \ldots, W_k$ such that $C_1 \cup C_2 \cup \cdots \cup C_k = V$, and for every vertex $v \in V$, $\sum_{i\,:\,v\in C_i} W_i \geq w(v)$.
The weight of a clique cover is defined as $\sum_{i=1}^k W_i$ and provides an upper bound on $\alpha_w(G)$.
This holds because the intersection of a clique and any IS of $G$ is either a single vertex or empty.
The objective then is to find a clique cover of small weight.
This can be done using an algorithm similar to the coloring method of Brelaz~\cite{brelazcoloring}.
However, this method can become computationally expensive since its running time is dependent on the maximum weight of the graph~\cite{warren2006combinatorial}.
Thus, we use a faster method to compute a weighted clique cover which is similar to the one used in Akiba and Iwata~\cite{akiba-tcs-2016}.

We begin by sorting the vertices in descending order of their weight (ties are broken by selecting the vertex with higher degree).
Next, we initiate an empty set of cliques $\mathcal{C}$.
We then iterate over the sorted vertices and search for the clique with maximum weight which it can be added to.
If there are no candidates for insertion, we insert a new single vertex clique to $\mathcal{C}$ and assign it the weight of the vertex.
Afterwards the vertex is marked as processed and we continue with the next one.
Computing a weighted clique cover using this algorithm has a linear running time independent of the maximum weight.
Thus, we are able to obtain a bound much faster.
However, this algorithm produces a higher weight clique cover than the method of Brelaz~\cite{brelazcoloring,akiba-tcs-2016}.

In addition to computing an upper bound, we also add an additional lower bound using a heuristic approach.
In particular, we run a modified version of the ILS by Andrade~\etal~\cite{andrade-2012} that is able to handle vertex weights for a fixed fraction of our total running time.
This lower bound is computed once after we apply our reductions initially and then again when splitting the search space on connected components.

\subsection{Connected Components.}
Solving the maximum weight independent set problem for a graph $G$ is equal to solving the problem for all $c$ connected components $G_1, \dots, G_{c}$ of $G$ and then combining the solution sets $\mathcal{I}_1, \dots, \mathcal{I}_{c}$ to form a solution $\mathcal{I}$ for $G$: $\mathcal{I} =  \bigcup_{i=1}^{c} \mathcal{I}_{i}$. We leverage this property by checking the connectivity of $G$ after each completed round of reduction applications.  If the graph disconnects due to branching or reductions then we apply our branch-and-reduce algorithm recursively to each of the connected components and combine their solutions afterwards. This technique can reduce the size of the branch-and-bound tree significantly on some instances.

\subsection{Branching.}
Our algorithm has to pick a branching order for the remaining vertices in the graph. 
Initially, vertices are sorted in non-decreasing order by degree, with ties broken by weight. 
Throughout the algorithm, the next vertex to be chosen is the highest vertex in the ordering.
This way our algorithm quickly eliminates the largest neighborhoods and makes the problem ``simpler''.

\section{Weighted Reduction Rules}
\label{sec:labelofchapterforoutlineofpaper_reductions}
\label{sec:new-reductions}
We now develop a comprehensive set of reduction rules for the maximum weight independent set problem.
We first introduce two \emph{meta} reductions, which we then use to instantiate many efficient reductions similar to already-known unweighted reductions. 

\begin{figure*}[t]
\centering
\includegraphics[scale=0.8]{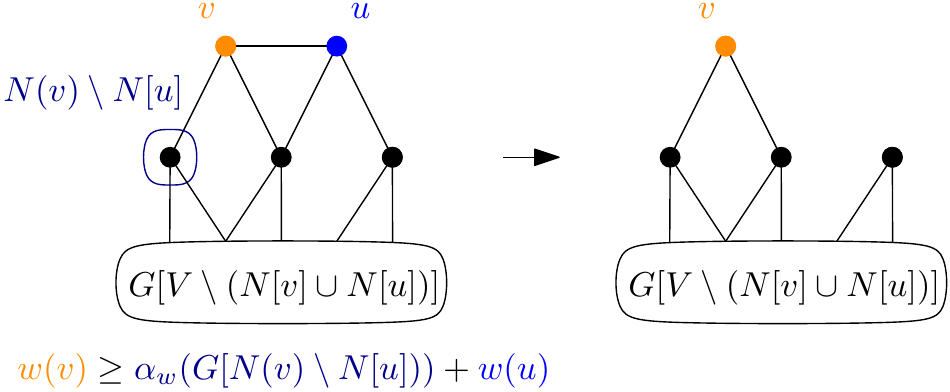} \includegraphics[scale=0.8]{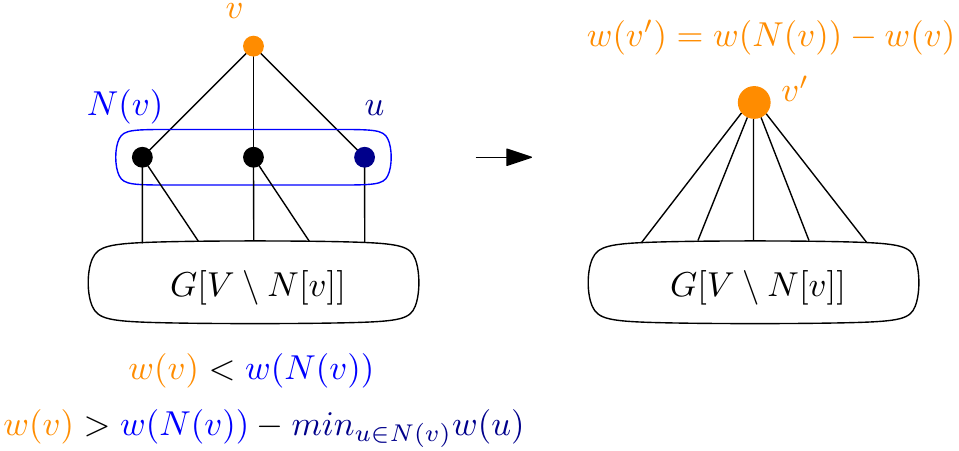}
\caption{Neighbor removal (left) and neighborhood folding (right)}
\label{fig:general_folding}
\label{fig:neighbor_removal}
\vspace*{-.5cm}
\end{figure*}

\subsection{Meta Reductions.}
\label{sec:general-reductions}
There are two operations that are commonly used in reductions: vertex removal and vertex folding. In the following reductions, we show general ways to detect when these operations can be applied in the neighborhood of a vertex.

\myparagraph{Neighbor Removal.}
In our first meta reduction, we show how to determine if a neighbor can be outright removed from the graph. We call this reduction the \emph{neighbor removal} reduction. (See Figure~\ref{fig:neighbor_removal}.)

\begin{reduction}[Neighbor Removal]
Let $v \in V$. For any $u\in N(v)$, if $\alpha_w(G[N(v) \setminus N[u]]) + w(u) \leq  w(v)$, then $u$ can be removed from $G$, as there is some MWIS of $G$ that excludes $u$, and $\alpha_w(G) = \alpha_w(G[V\setminus\{u\}])$.
\end{reduction}
\begin{proof}
Let $\I$ be an MWIS of $G$. We show by a cut-and-paste argument that if $u\in \I$ then there is another MWIS $\I'$ that contains $v$ instead. Let $u\in N(v)$, and suppose that $\alpha_w(G[N(v) \setminus N[u]]) + w(u) \leq  w(v)$. There are two cases, if $u$ is not in $\I$ then it is safe to remove. Otherwise, suppose $u\in \I$. Then $v\in N(u)$ is not in $\I$, and $w(\I\cap N(v)) = w(\I \cap (N(v)\setminus N[u])\cup\{u\}) = \alpha_w(G[N(v) \setminus N[u]]) + w(u) = w(v)$; otherwise we can swap $\I\cap N(v)$ for $v$ in $\I$ obtaining an independent set of larger weight. Thus $\I' = (\I \setminus N(v)) \cup \{v\}$ is an MWIS of $G$ excluding $u$ and $\alpha_w(G) = \alpha_w(G[V\setminus\{u\}])$. 
\end{proof}

\myparagraph{Neighborhood Folding.}
For our next meta reduction, we show a general condition for folding a vertex with its neighborhood. We first briefly describe the intuition behind the reduction. Consider $v$ and its neighborhood $N(v)$. If $N(v)$ has a unique independent set $\I_{N(v)}$ with weight larger than $w(v)$, then we only need to consider two independent sets: independent sets that contain $v$ or $\I_{N(v)}$. Otherwise, any other independent set in $N(v)$ can be swapped for $v$ and achieve higher overall weight. By folding $v$ with $\I_{N(v)}$, we can solve the remaining graph and then decide which of the two options will give an MWIS of the graph. (See Figure~\ref{fig:general_folding}.)
\begin{reduction}[Neighborhood Folding]
Let $v \in V$, and suppose that $N(v)$ is independent. If $w(N(v)) > w(v)$, but $w(N(v)) - \min_{u\in N(v)}\{w(u)\} < w(v)$, then fold $v$ and $N(v)$ into a new vertex $v'$ with weight $w(v') = w(N(v)) - w(v)$. Let $\I'$ be an MWIS of $G'$, then we construct an MWIS $\I$ of $G$ as follows: If $v'\in \I'$ then $\I = (\I'\setminus\{v'\}) \cup N(v)$, otherwise if $v\in \I'$ then $\I = \I' \cup \{v\}$. Furthermore, $\alpha_w(G) = \alpha_w(G') + w(v)$.
\end{reduction}
\begin{proof}
Proof can be found in Appendix~\ref{ommittedproofs}.
\end{proof}

However, these reductions require solving the MWIS problem on the neighborhood of a vertex, and therefore may be as expensive as computing an MWIS on the input graph. We next show how to use these meta reductions to develop efficient reductions.

\subsection{Efficient Weighted Reductions.}
We now construct new efficient reductions using the just defined meta reductions.
\myparagraph{Neighborhood Removal.}
In their HILS local search algorithm, Nogueria~\etal~\cite{hybrid-ils-2018} introduced the notion of a ``$(\omega,1)$-swap'', which swaps a vertex $v$ into a solution if its neighbors in the current solution $I$ have weight $w(N(v)\cap I) < w(v)$. This can be transformed into what we call the \emph{neighborhood removal reduction}.

\begin{reduction}[Neighorhood Removal]
For any $v\in V$, if $w(v)\geq w(N(v))$ then $v$ is in some MWIS of $G$. Let $G' = G[V\setminus N[v]]$ and $\alpha_w(G) = \alpha_w(G') + w(v)$.
\end{reduction}
\begin{proof}

\iffalse
We give a cut-and-paste argument. Suppose $w(v) \geq w(N(v))$, and let $\mathcal{I}$ be an MWIS of $G$. Either $v \in \mathcal{I}$, or not. If it is, then we are done. Otherwise, it must be that $w(N(v)) = w(v)$, otherwise, the independent set $\mathcal{I}'=(\mathcal{I}\setminus N(v)) \cup \{v\}$ has weight higher than $\mathcal{I}$. $\mathcal{I}'$ has the same weight as $\mathcal{I}$ and is an MWIS containing $v$.
\else
Since $w(N(v)) \leq w(v)$, $\forall u\in N(v)$ we have that
\[\alpha_w(G[N(v)\cap N(u)]) + w(u) \leq w(N(v)) \leq w(v).\] Then we can remove all $u\in N(v)$ and are left with $v$ in its own component. Calling this graph $G'$, we have that $v$ is in some MWIS and $\alpha_w(G) = \alpha_w(G') + w(v)$.
\fi{}
\end{proof}
For the remaining reductions, we assume that the neighborhood removal reduction has already been applied. Thus, $\forall v\in V$, $w(v) < w(N(v))$.

\myparagraph{Weighted Isolated Vertex Removal.} Similar to the (unweighted) isolated vertex removal reduction, we now argue that an isolated vertex is in some MWIS---if it has highest weight in its clique.
\begin{reduction}[Isolated Vertex Removal.]
Let $v\in V$ be isolated and $w(v)\geq \max_{u\in N(v)}w(u)$. Then $v$ is in some MWIS of $G$. Let $G' = G[V\setminus N[v]]$ and $\alpha_w(G) = \alpha_w(G') + w(v)$.
\end{reduction}
\begin{proof}
\iffalse
By a cut-and-paste argument, let $\mathcal{I}\subseteq V$ be an MWIS of $G$. Then either $v$ is in $\mathcal{I}$ and we are done, or not. If $v\notin\mathcal{I}$, then there must be a neighbor $u\in \mathcal{I}\cap N(v)$ such that $w(u) = w(v)$. Otherwise the independent set $\mathcal{I}' = (\mathcal{I} \setminus N(v)) \cup \{v\}$ has higher weight than $\mathcal{I}$. $\mathcal{I}'$ is an MWIS containing $v$.
\else
Since $N(v)$ is a clique, $\forall u\in N(v)$ we have that
\[\alpha_w(G[N(v)\cap N(u)]) \leq \alpha_w(N(v)) = \max_{u\in N(v)}\{w(u)\} \leq w(v).\]
Similar to neighborhood removal, remove all $u\in N(v)$ producing $G'$ and $\alpha_w(G) = \alpha_w(G') + w(v)$.
\fi 
\end{proof}

\vspace*{-.25cm}
\myparagraph{Isolated Weight Transfer.} Given its weight restriction, the weighted isolated vertex removal reduction may be ineffective. We therefore introduce a reduction that supports more liberal vertex removal.  
\begin{figure}[t]
\centering
\includegraphics[scale=0.8]{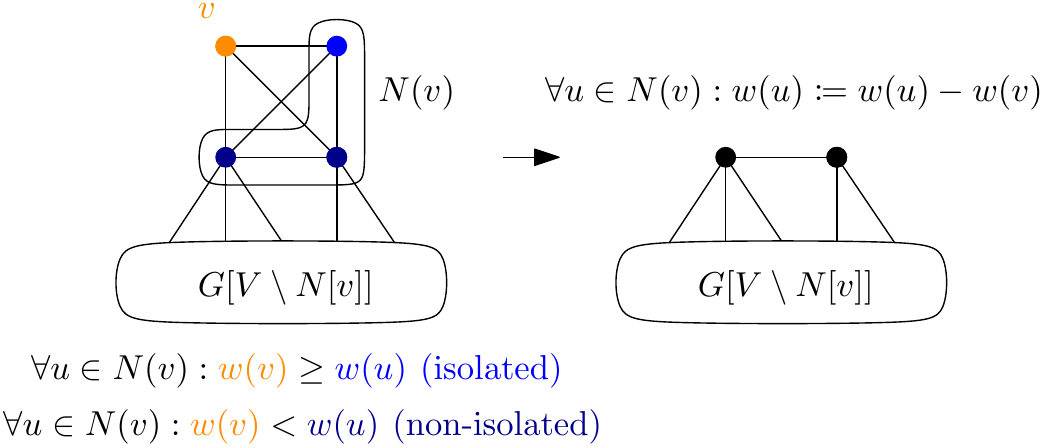}
\caption{Isolated weight transfer}
\label{fig:weighted_clique}
\end{figure}

\begin{reduction}[Isolated weight transfer]
Let $v\in V$ be isolated, and suppose that the set of isolated vertices $S(v)\subseteq N(v)$ is such that $\forall u\in S(v)$, $w(v) \geq w(u)$. We
\begin{enumerate}[(i)]
\item remove all $u\in N(v)$ such that $w(u)\leq w(v)$, and let the remaining neighbors be denoted by $N'(v)$,
\item remove $v$ and $\forall x\in N'(v)$ set its new weight to $w'(x) = w(x) - w(v)$, and
\end{enumerate}
let the resulting graph be denoted by $G'$. Then $\alpha_w(G) = w(v) + \alpha_w(G')$ and an MWIS $\I$ of $G$ can be constructed from an MWIS $\I'$ of $G'$ as follows: if $\I' \cap N'(v) = \emptyset$ then $\I = \I'\cup\{v\}$, otherwise $\I = \I'$.
\end{reduction}
\begin{proof}
Proof can be found in Appendix~\ref{ommittedproofs}.%
\end{proof}

\vspace*{-.25cm}
\myparagraph{Weighted Vertex Folding.} Similar to the unweighted vertex folding reduction, we show that we can fold vertices with two non-adjacent neighbors---however, not all weight configurations permit this.
\begin{figure}[t]
\centering
\includegraphics[scale=0.8]{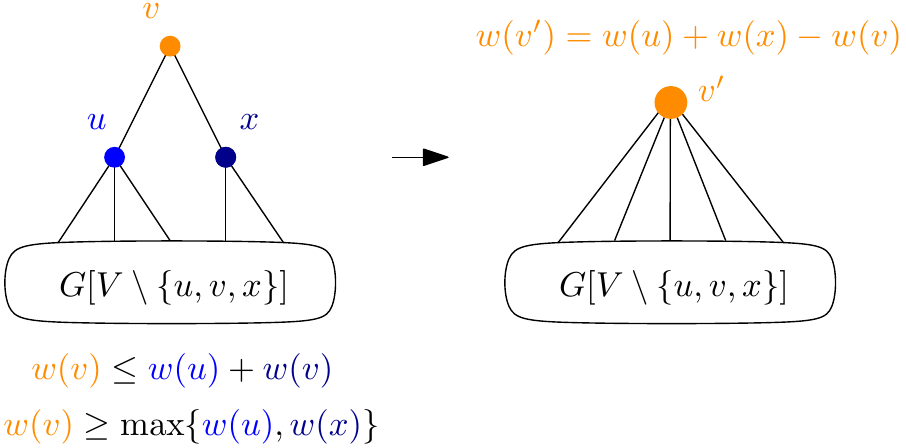}
\caption{Weighted vertex folding}
\label{fig:weighted_folding}
\vspace*{-.25cm}
\end{figure}
\begin{reduction}[Vertex Folding]
Let $v\in V$ have $d(v) = 2$, such that $v$'s neighbors $u$, $x$ are not adjacent.
If $w(v) < w(u) + w(x)$ but $w(v)\geq \max\{w(u),w(x)\}$, then we fold $v,u,x$ into vertex $v'$ with weight $w(v') = w(u) + w(x) - w(v)$ forming a new graph $G'$. Then $\alpha_w(G) = \alpha_w(G') + w(v)$. Let $\mathcal{I}'$ be an MWIS of $G'$. If $v'\in \mathcal{I}'$ then $\I = (\I'\setminus\{v'\})\cup\{u,x\}$ is an MWIS of $G$. Otherwise, $\I = \I' \cup \{v\}$ is an MWIS of $G$.
\end{reduction}
\begin{proof}
\iffalse

First note that after folding, the following graphs are identical: $G[V\setminus N[\{u,x\}]] = G'[V'\setminus N_{G'}[\{v'\}]]$ and $G'[V'\setminus \{v'\}] = G[V\setminus N[v]]$.
Let $\I'$ be an MWIS of $G'$. Then we have two cases.

\noindent \emph{Case 1 ($v'\in\I'$):}
Suppose that $v'\in\I'$. We show that $w(u) + w(x) + \alpha_w(G[V\setminus N[\{u,x\}]]) \geq w(v) + \alpha_w(G[V\setminus N[v]])$, which shows that $u$ and $x$ are together in some MWIS of $G$.

\noindent Since $v'\in\I'$, we have that
\begin{align*}
w(v) + \alpha_w(G') &= w(v) + w(v') + \alpha_w(G'[V'\setminus N_{G'}[v']])\\
                    &= w(v) + w(u) + w(x) - w(v) \\
                    &\phantom{= w(v) + w(u)\text{ }} + \alpha_w(G'[V'\setminus N_{G'}[v']])\\
                    &= w(u) + w(x) + \alpha_w(G[V\setminus N[\{u,x\}]]).
\end{align*}

\noindent But since $\I'$ is an MWIS of $G'$, we have that
\begin{align*}
w(v) + \alpha_w(G') &\geq w(v) + \alpha_w(G'[V'\setminus \{v'\}]) \\
                    &= w(v) + \alpha_w(G[V\setminus\{u,x\}]) \\
                    &= w(v) + \alpha_w(G[V\setminus N[v]]).
\end{align*}

\noindent Thus,
$w(u) + w(x) + \alpha_w(G[V\setminus N[\{u,x\}]]) \geq w(v) + \alpha_w(G[V\setminus N[v]])$ and
$u$ and $x$ are together in some MWIS of $G$.
Furthermore, 
we have that
\begin{align*}
\alpha_w(G) &= w(u) + w(x) + \alpha_w(G[V\setminus N[\{u,x\}]]) \\ &= \alpha_w(G') + w(v).
\end{align*}

\noindent \emph{Case 2: ($v'\notin\I'$):}
Suppose that $v'\notin\I'$. We show that $w(v) + \alpha_w(G[V\setminus N[v]]) \geq w(u) + w(x) + \alpha_w(G[V\setminus N[\{u,x\}]])$, which shows that $v$ is in some MWIS of $G$.

\noindent Since $v'\notin\I'$, we have that
\begin{align*}
w(v) + \alpha_w(G') &= w(v) + \alpha_w(G'[V'\setminus \{v'\}])\\
                    &= w(v) + \alpha_w(G[V\setminus N[v]])
\end{align*}

\noindent But since $\I'$ is an MWIS of $G'$, we have that
\begin{align*}
w(v) + \alpha_w(G') &\geq w(v) + w(v') + \alpha_w(G'[V'\setminus N_{G'}[v']]) \\
                    &= w(v) + w(u) + w(x) - w(v) \\
                    &\phantom{= w(v) + w(u)\text{ } }+ \alpha_w(G[V\setminus N[\{u,x\}]]) \\
                    &= w(u) + w(x) + \alpha_w(G[V\setminus N[\{u,x\}]]).
\end{align*}

\noindent Thus,
$w(v) + \alpha_w(G[V\setminus N[v]]) \geq w(u) + w(x) + \alpha_w(G[V\setminus N[\{u,x\}]])$ and
$v$ is in some MWIS of $G$.
Lastly,
\begin{align*}
\alpha_w(G) &= w(v) + \alpha_w(G[V\setminus N[v]]) \\ &= \alpha_w(G') + w(v).
\end{align*}
\else
Apply neighborhood folding to $v$.
\fi{} 
\end{proof}

\vspace*{-.25cm}
\myparagraph{Weighted Twin.}
\begin{figure}[t]
\centering
\includegraphics[scale=0.8]{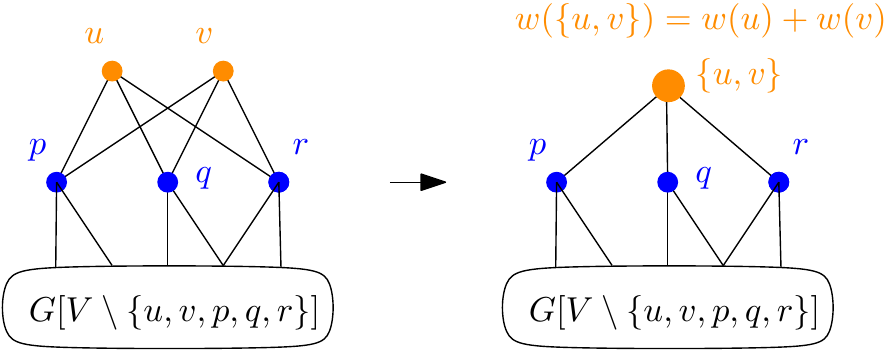}
\caption{Illustrating proof of weighted twin reduction}
\label{fig:weighted_twin}
\end{figure}
The twin reduction, as described by Akiba and Iwata~\cite{akiba-tcs-2016} for the unweighted case, works for twins with $3$ common neighbors. We describe our variant in the same terms, but note that the reduction supports an arbitrary number of common neighbors. 

\begin{reduction}[Twin]
Let vertices $u$ and $v$ \ifFull be twins with \else have \fi{} independent neighborhoods $N(u) = N(v) = \{p,q,r\}$. 
We have two cases:
\begin{enumerate}[(i)]
\item If $w(\{u,v\}) \geq w(\{p,q,r\})$, then $u$ and $v$ are in some MWIS of $G$. Let $G' = G[V\setminus N[\{u,v\}]]$.
\item If $w(\{u,v\}) < w(\{p,q,r\})$, but $w(\{u,v\}) > w(\{p,q,r\}) - \min_{x\in\{p,q,r\}} w(x)$, then we can fold $u, v, p, q, r$ into a new vertex $v'$ with weight $w(v') = w(\{p,q,r\}) - w(\{u,v\})$ and call this graph $G'$. Let $I'$ be an MWIS of $G'$. Then we construct an MWIS $\I$ of $G$ as follows: if $v'\in \I'$ then $\I = (\I'\setminus \{v'\})\cup \{p,q,r\}$, if $v'\notin \I'$ then $\I = \I' \cup \{u,v\}$.
\end{enumerate}
Furthermore, $\alpha_w(G) = \alpha_w(G') + w(\{u,v\})$.
\end{reduction}
\begin{proof}
Just as in the unweighted case, either $u$ and $v$ are simultaneously in an MWIS or some subset of $p$, $q$, $r$ is in. 
First fold $u$ and $v$ into a new vertex $\{u,v\}$ with weight $w(\{u,v\})$. To show (i), apply the neighborhood reduction to vertex $\{u,v\}$. For (ii), since $N(\{u,v\})$ is independent, we 
apply the neighborhood folding reduction to $\{u,v\}$, giving the claimed result.%
\end{proof}

If $p,q,r$ are not independent, further reductions are possible; however, introducing a comprehensive list is not illuminating. Instead, we can simply let meta reductions reduce as appropriate.

\myparagraph{Weighted Domination.} Lastly, we give a weighted variant of the domination reduction.
\begin{figure}[t]
\centering
\includegraphics[scale=0.8]{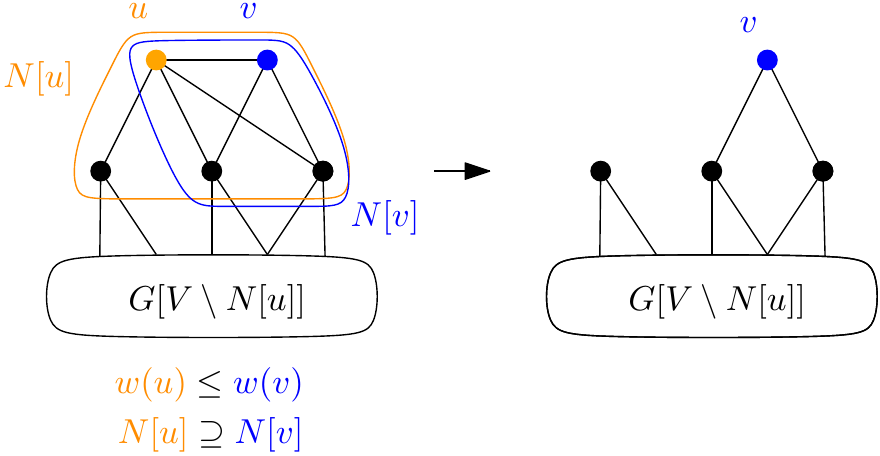}
\caption{Weighted domination}
\label{fig:weighted_domination}

\vspace*{-.5cm}
\end{figure}

\begin{reduction}[Domination]
Let $u,v\in V$ be vertices such that $N[u]\supseteq N[v]$ (i.e., $u$ dominates $v$). If $w(u)\leq w(v)$, there is an MWIS in $G$ that excludes $u$ and $\alpha_w(G) = \alpha_w(G[V\setminus \{u\}])$. Therefore, $u$ can be removed from the graph.
\end{reduction}
\begin{proof}
We show by a cut-and-paste argument that there is an MWIS of $G$ excluding $u$. Let $\I$ be an MWIS of $G$. If $u$ is not in $\I$ then we are done. Otherwise, suppose $u\in \I$. Then it must be the case that $w(v) = w(u)$, otherwise $\I' = (\I\setminus\{u\}) \cup \{v\}$ is an independent set with weight larger than $\I$. Thus, $\I'$ is an MWIS of $G$ excluding $u$, and $\alpha_w(G) = \alpha_w(G[V\setminus \{u\}])$.
\end{proof}

\section{Experimental Evaluation}
\label{sec:experiments}
We now compare the performance of our branch-and-reduce algorithm to existing state-of-the-art algorithms on large real-world graphs.
Furthermore, we examine how our reduction rules can drastically improve the quality of existing heuristic approaches.

\vspace*{-.25cm}
\subsection{Methodology and Setup.}
All of our experiments were run on a machine with four Octa-Core Intel Xeon E5-4640 processors running at $2.4$ GHz, $512$ GB of main memory, $420$ MB L3-Cache and $48256$ KB L2-Cache.
The machine runs Ubuntu 14.04.3 and Linux kernel version 3.13.0-77.
All algorithms were implemented in \Cminus11 and compiled with \GG~version 4.8.4 with optimization flag \texttt{-O3}.
Each algorithm was run sequentially for a total of $1000$ seconds\footnote{Results with more than 1000 seconds are due to initial kernelization taking longer than the time limit.}. 
We present two kinds of data: (1) the best solution found by each algorithm and the time (in seconds) required to obtain it, (2) \emph{convergence plots}, which show how the solution quality changes over time.
In particular, whenever an algorithm finds a new best independent set $S$ at time $t$, it reports a tuple ($t$, $|S|$)\footnote{For the convergence plots of the heuristic algorithms we use the maximum values over five runs with varying random seeds}.
\begin{table*}[ht]
\scriptsize
\centering
\setlength{\tabcolsep}{0.5ex}
\begin{tabular}{l|r|r r|r r|r r} 
Graph & $|V|$ & $t_\text{max}$ & $w_\text{max}$ & $t_\text{max}$ & $w_\text{max}$ & $t_\text{max}$ & $w_\text{max}$ \\ 
\hline 
\rule{0pt}{3ex}OSM networks & & \multicolumn{2}{c|}{DynWVC1} & \multicolumn{2}{c|}{HILS} & \multicolumn{2}{c}{\BRD} \\ 
\hline 
\Id{\detokenize{alabama-AM3}} & \numprint{3504} & \numprint{464.02} & \numprint{185527} & \numprint{0.73} & \textbf{\numprint{185744}} & \numprint{15.79} & \numprint{185707} \\ 
\rowcolor{lightergray} \Id{\detokenize{florida-AM2}} & \numprint{1254} & \numprint{1.14} & \textbf{\numprint{230595}} & \numprint{0.04} & \textbf{\numprint{230595}} & \numprint{0.03} & \textbf{\numprint{230595}} \\ 
\Id{\detokenize{georgia-AM3}} & \numprint{1680} & \numprint{0.88} & \textbf{\numprint{222652}} & \numprint{0.05} & \textbf{\numprint{222652}} & \numprint{4.88} & \numprint{214918} \\ 
\Id{\detokenize{kansas-AM3}} & \numprint{2732} & \numprint{46.87} & \textbf{\numprint{87976}} & \numprint{0.84} & \textbf{\numprint{87976}} & \numprint{11.35} & \numprint{87925} \\ 
\rowcolor{lightergray} \Id{\detokenize{maryland-AM3}} & \numprint{1018} & \numprint{1.34} & \textbf{\numprint{45496}} & \numprint{0.02} & \textbf{\numprint{45496}} & \numprint{3.34} & \textbf{\numprint{45496}} \\ 
\Id{\detokenize{massachusetts-AM3}} & \numprint{3703} & \numprint{435.31} & \numprint{145863} & \numprint{2.73} & \textbf{\numprint{145866}} & \numprint{12.87} & \numprint{145617} \\ 
\rowcolor{lightergray} \Id{\detokenize{utah-AM3}} & \numprint{1339} & \numprint{136.15} & \numprint{98802} & \numprint{0.08} & \textbf{\numprint{98847}} & \numprint{64.04} & \textbf{\numprint{98847}} \\ 
\Id{\detokenize{vermont-AM3}} & \numprint{3436} & \numprint{119.63} & \numprint{63234} & \numprint{0.95} & \textbf{\numprint{63302}} & \numprint{95.81} & \numprint{55584} \\ 
\hline 
\rule{0pt}{3ex}Solved instances & & \multicolumn{2}{r|}{} & \multicolumn{2}{r|}{} & \multicolumn{2}{r}{44.12\% (15/34)  } \\ 
Optimal weight & & \multicolumn{2}{r|}{60.00\% (9/15)  }  & \multicolumn{2}{r|}{100.00\% (15/15)  }  & \multicolumn{2}{r}{} \\ 
\rule{0pt}{4ex}SNAP networks & & \multicolumn{2}{c|}{DynWVC2} & \multicolumn{2}{c|}{HILS} & \multicolumn{2}{c}{\BRF} \\ 
\hline 
\Id{\detokenize{as-skitter}} & \numprint{1696415} & \numprint{576.93} & \numprint{123105765} & \numprint{998.75} & \numprint{122539706} & \numprint{746.93} & \textbf{\numprint{123904741}} \\ 
\rowcolor{lightergray} \Id{\detokenize{ca-AstroPh}} & \numprint{18772} & \numprint{108.35} & \numprint{796535} & \numprint{46.76} & \textbf{\numprint{796556}} & \numprint{0.03} & \textbf{\numprint{796556}} \\ 
\rowcolor{lightergray} \Id{\detokenize{email-EuAll}} & \numprint{265214} & \numprint{179.26} & \textbf{\numprint{25330331}} & \numprint{501.09} & \textbf{\numprint{25330331}} & \numprint{0.19} & \textbf{\numprint{25330331}} \\ 
\rowcolor{lightergray} \Id{\detokenize{p2p-Gnutella08}} & \numprint{6301} & \numprint{0.19} & \textbf{\numprint{435893}} & \numprint{0.25} & \textbf{\numprint{435893}} & \numprint{0.01} & \textbf{\numprint{435893}} \\ 
\rowcolor{lightergray} \Id{\detokenize{roadNet-TX}} & \numprint{1379917} & \numprint{1000.78} & \numprint{77525099} & \numprint{1697.13} & \numprint{76366577} & \numprint{33.49} & \textbf{\numprint{78606965}} \\ 
\Id{\detokenize{soc-LiveJournal1}} & \numprint{4847571} & \numprint{1001.23} & \numprint{277824322} & \numprint{12437.50} & \numprint{280559036} & \numprint{270.96} & \textbf{\numprint{283948671}} \\ 
\rowcolor{lightergray} \Id{\detokenize{web-Google}} & \numprint{875713} & \numprint{683.63} & \numprint{56190870} & \numprint{994.58} & \numprint{55954155} & \numprint{3.16} & \textbf{\numprint{56313384}} \\ 
\rowcolor{lightergray} \Id{\detokenize{wiki-Talk}} & \numprint{2394385} & \numprint{991.31} & \numprint{235874419} & \numprint{996.02} & \numprint{235852509} & \numprint{3.36} & \textbf{\numprint{235875181}} \\ 
\hline 
\rule{0pt}{3ex}Solved instances & & \multicolumn{2}{r|}{} & \multicolumn{2}{r|}{} & \multicolumn{2}{r}{80.65\% (25/31)  } \\ 
Optimal weight & & \multicolumn{2}{r|}{28.00\% (7/25)  }  & \multicolumn{2}{r|}{68.00\% (17/25)  }  & \multicolumn{2}{r}{} \\ 
\end{tabular} 

\caption{Best solution found by each algorithm and time (in seconds) required to compute it. The global best solution is highlighted in \textbf{bold}. Rows are highlighted in gray if B~\&~R is able to find an exact solution.
}
\label{tab:local_rt}
\vspace*{-.5cm}
\end{table*}

\myparagraph{Algorithms Compared.}
We use two different variants of our branch-and-reduce algorithm.
The first variant, called \BRF, uses our \emph{full set} of reductions each time we branch.
The second variant, called \BRD, omits the more costly reductions and also terminates the execution of the remaining reductions faster than \BRF.
In particular, this configuration completely omits the weighted critical set reductions from both the initialization and recursion.
Additionally, we also omit the weighted clique reduction from the first reduction call and use a faster version that only considers triangles during recursion.
Finally, we do not use the generalized neighborhood folding during recursion. 
This configuration find solutions more quickly on dense graphs.

We also include the state-of-the-art heuristics \emph{HILS} by Nogueria~\etal~\cite{hybrid-ils-2018} and both versions of \emph{DynWVC} by Cai~\etal~\cite{cai-dynwvc} (see Section~\ref{sec:related_work} for a short explanation of these algorithms).
Finally, we do not include any other exact algorithms (\eg~\cite{butenko-trukhanov,warren2006combinatorial}) as their code is not available.
Also note that these exact algorithms are either not tested in the weighted case~\cite{butenko-trukhanov} or the largest instances reported consist of a few hundred vertices~\cite{warren2006combinatorial}.

To further evaluate the impact of reductions on existing algorithms, we also propose combinations of the heuristic approaches with reductions (\emph{Red~+~HILS} and \emph{Red~+~DynWVC}).
We do so by first computing a kernel graph using our set of reductions and then run the existing algorithms on the resulting graph.

\myparagraph{Instances.}
We test all algorithms on a large corpus of sparse data sets.
For this purpose, we include a set of real-world conflict graphs obtained from OpenStreetMap \cite{OSMWEB} files of North America, according to the method described by Barth~\etal~\cite{barth-2016}.
More specifically, these graphs are generated by identifying map labels with vertices that have a weight corresponding to their importance.
Edges are then inserted between vertices if their labels overlap each other.
Conflict graphs can also be used in a dynamic setting by associating vertices with intervals that correspond to the time they are displayed.
Furthermore, solving the MWIS problem on these graphs eliminates label conflicts and maximizes the importance of displayed labels.
Finally, different activity models (AM1, AM2 and AM3) are used to generate different conflict graphs.
The instances we use for our experiments are the same ones used by Cai~\etal~\cite{cai-dynwvc}.
We omit all instances with less than $1000$ vertices from our experiments, as these are easy to solve and our focus is on large scale networks~\cite{cai-dynwvc}.

In addition to the OSM networks, we also include collaboration networks, communication networks, additional road networks, social networks, peer-to-peer networks, and Web crawl graphs from the Stanford Large Network Dataset Repository~\cite{snapnets} (SNAP).

These networks are popular benchmark instances commonly used for the maximum independent set problem~\cite{akiba-tcs-2016,dahlum2016,redumis-2017}.
However, all SNAP instances are unweighted and comparable weighted instances are very scarce.
Therefore, a common approach in literature is to assign vertex weights uniformly at random from a fixed size interval~\cite{cai-dynwvc,li2017efficient}.
To keep our results in line with existing work, we thus decided to select vertex weights uniformly at random from $[1,200]$.
Basic properties of our benchmark instances can be found in Table~\ref{tab:props}. 

\subsection{Comparison with State-of-the-Art.}
\label{sec:sota}
A representative sample of our experimental results for the OSM and SNAP networks is presented in Table~\ref{tab:local_rt}.
For a full overview of all instances, we refer to Table~\ref{tab:osm_local_rt} (OSM) and Table~\ref{tab:snap_local_rt} (SNAP) respectively. 
For each instance, we list the best solution computed by each algorithm $w_\text{Algo}$ and the time in seconds required to find it $t_\text{Algo}$.
For each data set, we highlight the best solution found across all algorithms in \textbf{bold}.
Additionally, if any version of our algorithm is able to find an exact solution, the corresponding row is highlighted in gray.
Finally, recall that our algorithm computes a solution on unsolved instances once the time-limit is reached by additionally running a greedy algorithm as post-processing.

Examining the OSM graphs, B~\&~R is able to solve 15 out of the 34 instances we tested.
However, HILS is also able to compute a solution with the same weight on all of these instance.
Furthermore, HILS obtains a higher or similar quality solution than both versions of DynWVC and B~\&~R for all remaining unsolved instances.
Overall, HILS is able to find the best solution on all OSM instances that we tested.
Additionally, on most of these instances it does so significantly faster than all of its competitors.
Note though, that neither HILS nor DynWVC provide any optimality guarantees (in contrast to B~\&~R).

Looking at both versions of DynWVC, we see that DynWVC1 performs better than DynWVC2, which is also reported by Cai~\etal~\cite{cai-dynwvc}.
Comparing both variants of our branch-and-reduce algorithm, we see that they are able to solve the same instances.
Nonetheless, \BRD~is able to compute better solutions on roughly half of the remaining instances.
Additionally, it almost always requires significantly less time to achieve its maximum compared to \BRF.

\begin{figure*}[t]
\vspace*{-0.5cm}
\centering
\includegraphics[scale=0.45]{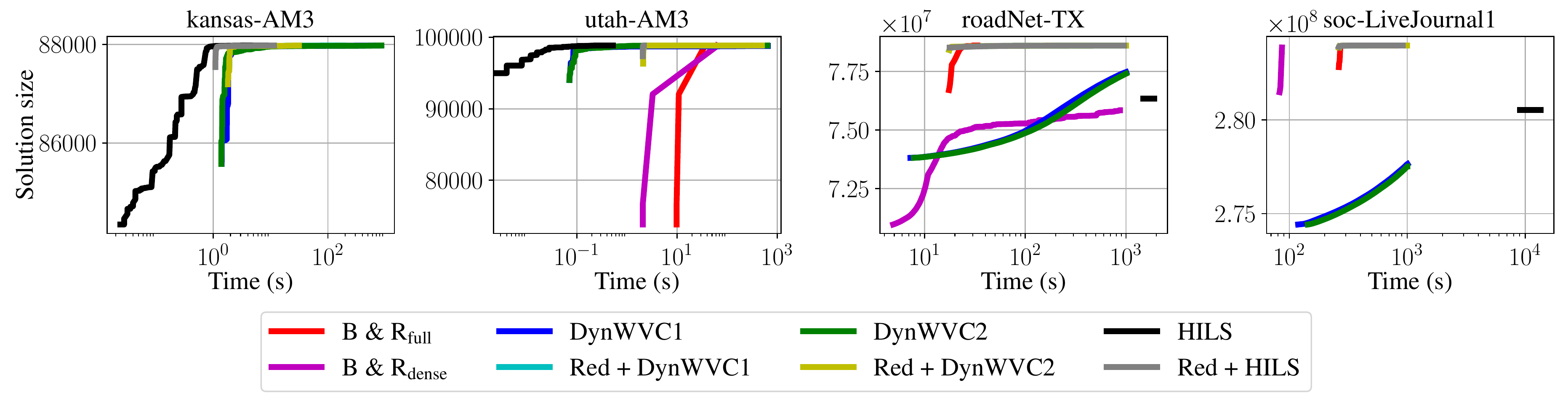}
\vspace*{-.25cm}
\caption{Solution quality over time for two OSM instances (left) and two SNAP instances (right).}
\label{fig:convergence}
\end{figure*}

\begin{table*}[ht]
\scriptsize
\centering
\setlength{\tabcolsep}{0.5ex}
\begin{tabular}{l|r r|r r r|r r|r r r|r r} 
Graph & $t_\text{max}$ & $w_\text{max}$ & $t_\text{max}$ & $w_\text{max}$ & $S_\text{base}$ & $t_\text{max}$ & $w_\text{max}$ & $t_\text{max}$ & $w_\text{max}$ & $S_\text{base}$ & $t_\text{max}$ & $w_\text{max}$ \\ 
\hline 
\rule{0pt}{4ex}OSM instances & \multicolumn{2}{c|}{DynWVC1} & \multicolumn{3}{c|}{Red+DynWVC1} & \multicolumn{2}{c|}{HILS} & \multicolumn{3}{c|}{Red + HILS} & \multicolumn{2}{c}{\BRD} \\ 
\hline 
\Id{\detokenize{alabama-AM3}} & \numprint{464.02} & \numprint{185527} & \numprint{370.80} & \numprint{185727} & \textcolor{darkgreen}{\numprint{1.25}} & \numprint{0.73} & \textbf{\numprint{185744}} & \numprint{4.05} & \textbf{\numprint{185744}} & \textcolor{darkred}{\numprint{0.18}} & \numprint{15.79} & \numprint{185707} \\ 
\rowcolor{lightergray} \Id{\detokenize{florida-AM2}} & \numprint{1.14} & \textbf{\numprint{230595}} & \numprint{0.03} & \textbf{\numprint{230595}} & \textcolor{darkgreen}{\numprint{44.19}} & \numprint{0.04} & \textbf{\numprint{230595}} & \numprint{0.03} & \textbf{\numprint{230595}} & \textcolor{darkgreen}{\numprint{1.75}} & \numprint{0.03} & \textbf{\numprint{230595}} \\ 
\Id{\detokenize{georgia-AM3}} & \numprint{0.88} & \textbf{\numprint{222652}} & \numprint{2.64} & \textbf{\numprint{222652}} & \textcolor{darkred}{\numprint{0.33}} & \numprint{0.05} & \textbf{\numprint{222652}} & \numprint{2.43} & \textbf{\numprint{222652}} & \textcolor{darkred}{\numprint{0.02}} & \numprint{4.88} & \numprint{214918} \\ 
\Id{\detokenize{kansas-AM3}} & \numprint{46.87} & \textbf{\numprint{87976}} & \numprint{13.59} & \textbf{\numprint{87976}} & \textcolor{darkgreen}{\numprint{3.45}} & \numprint{0.84} & \textbf{\numprint{87976}} & \numprint{2.06} & \textbf{\numprint{87976}} & \textcolor{darkred}{\numprint{0.41}} & \numprint{11.35} & \numprint{87925} \\ 
\rowcolor{lightergray} \Id{\detokenize{maryland-AM3}} & \numprint{1.34} & \textbf{\numprint{45496}} & \numprint{2.07} & \textbf{\numprint{45496}} & \textcolor{darkred}{\numprint{0.65}} & \numprint{0.02} & \textbf{\numprint{45496}} & \numprint{2.07} & \textbf{\numprint{45496}} & \textcolor{darkred}{\numprint{0.01}} & \numprint{3.34} & \textbf{\numprint{45496}} \\ 
\Id{\detokenize{massachusetts-AM3}} & \numprint{435.31} & \numprint{145863} & \numprint{10.68} & \textbf{\numprint{145866}} & \textcolor{darkgreen}{\numprint{40.75}} & \numprint{2.73} & \textbf{\numprint{145866}} & \numprint{2.92} & \textbf{\numprint{145866}} & \textcolor{darkred}{\numprint{0.93}} & \numprint{12.87} & \numprint{145617} \\ 
\rowcolor{lightergray} \Id{\detokenize{utah-AM3}} & \numprint{136.15} & \numprint{98802} & \numprint{168.07} & \textbf{\numprint{98847}} & \textcolor{darkred}{\numprint{0.81}} & \numprint{0.08} & \textbf{\numprint{98847}} & \numprint{2.10} & \textbf{\numprint{98847}} & \textcolor{darkred}{\numprint{0.04}} & \numprint{64.04} & \textbf{\numprint{98847}} \\ 
\Id{\detokenize{vermont-AM3}} & \numprint{119.63} & \numprint{63234} & \numprint{62.85} & \numprint{63280} & \textcolor{darkgreen}{\numprint{1.90}} & \numprint{0.95} & \numprint{63302} & \numprint{2.95} & \textbf{\numprint{63312}} & \textcolor{darkred}{\numprint{0.32}} & \numprint{95.81} & \numprint{55584} \\ 
\hline 
\rule{0pt}{3ex}Solved instances & \multicolumn{2}{r|}{} & \multicolumn{3}{r|}{} & \multicolumn{2}{r|}{} & \multicolumn{3}{r|}{} & \multicolumn{2}{r}{44.12\% (15/34)  } \\ 
Optimal weight & \multicolumn{2}{r|}{60.00\% (9/15)  }  & \multicolumn{3}{r|}{93.33\% (14/15)  }  & \multicolumn{2}{r|}{100.00\% (15/15)  }  & \multicolumn{3}{r|}{100.00\% (15/15)  } & \multicolumn{2}{r}{} \\ 
\rule{0pt}{4ex}SNAP instances & \multicolumn{2}{c|}{DynWVC2} & \multicolumn{3}{c|}{Red+DynWVC2} & \multicolumn{2}{c|}{HILS} & \multicolumn{3}{c|}{Red + HILS} & \multicolumn{2}{c}{\BRD} \\ 
\hline 
\Id{\detokenize{as-skitter}} & \numprint{576.93} & \numprint{123105765} & \numprint{85.60} & \numprint{123995808} & \textcolor{darkgreen}{\numprint{6.74}} & \numprint{998.75} & \numprint{122539706} & \numprint{845.70} & \textbf{\numprint{123996322}} & \textcolor{darkgreen}{\numprint{1.18}} & \numprint{746.93} & \numprint{123904741} \\ 
\rowcolor{lightergray} \Id{\detokenize{ca-AstroPh}} & \numprint{108.35} & \numprint{796535} & \numprint{0.02} & \textbf{\numprint{796556}} & \textcolor{darkgreen}{\numprint{4962.17}} & \numprint{46.76} & \textbf{\numprint{796556}} & \numprint{0.02} & \textbf{\numprint{796556}} & \textcolor{darkgreen}{\numprint{2142.48}} & \numprint{0.03} & \textbf{\numprint{796556}} \\ 
\rowcolor{lightergray} \Id{\detokenize{email-EuAll}} & \numprint{179.26} & \textbf{\numprint{25330331}} & \numprint{0.12} & \textbf{\numprint{25330331}} & \textcolor{darkgreen}{\numprint{1548.08}} & \numprint{501.09} & \textbf{\numprint{25330331}} & \numprint{0.12} & \textbf{\numprint{25330331}} & \textcolor{darkgreen}{\numprint{4327.82}} & \numprint{0.19} & \textbf{\numprint{25330331}} \\ 
\rowcolor{lightergray} \Id{\detokenize{p2p-Gnutella08}} & \numprint{0.19} & \textbf{\numprint{435893}} & \numprint{0.00} & \textbf{\numprint{435893}} & \textcolor{darkgreen}{\numprint{46.98}} & \numprint{0.25} & \textbf{\numprint{435893}} & \numprint{0.00} & \textbf{\numprint{435893}} & \textcolor{darkgreen}{\numprint{63.80}} & \numprint{0.01} & \textbf{\numprint{435893}} \\ 
\rowcolor{lightergray} \Id{\detokenize{roadNet-TX}} & \numprint{1000.78} & \numprint{77525099} & \numprint{771.05} & \numprint{78601813} & \textcolor{darkgreen}{\numprint{1.30}} & \numprint{1697.13} & \numprint{76366577} & \numprint{946.32} & \numprint{78602984} & \textcolor{darkgreen}{\numprint{1.79}} & \numprint{33.49} & \textbf{\numprint{78606965}} \\ 
\Id{\detokenize{soc-LiveJournal1}} & \numprint{1001.23} & \numprint{277824322} & \numprint{996.68} & \numprint{283973997} & \textcolor{darkgreen}{\numprint{1.00}} & \numprint{12437.50} & \numprint{280559036} & \numprint{761.51} & \textbf{\numprint{283975036}} & \textcolor{darkgreen}{\numprint{16.33}} & \numprint{270.96} & \numprint{283948671} \\ 
\rowcolor{lightergray} \Id{\detokenize{web-Google}} & \numprint{683.63} & \numprint{56190870} & \numprint{3.30} & \numprint{56313349} & \textcolor{darkgreen}{\numprint{207.26}} & \numprint{994.58} & \numprint{55954155} & \numprint{3.01} & \textbf{\numprint{56313384}} & \textcolor{darkgreen}{\numprint{330.28}} & \numprint{3.16} & \textbf{\numprint{56313384}} \\ 
\rowcolor{lightergray} \Id{\detokenize{wiki-Talk}} & \numprint{991.31} & \numprint{235874419} & \numprint{2.30} & \textbf{\numprint{235875181}} & \textcolor{darkgreen}{\numprint{430.22}} & \numprint{996.02} & \numprint{235852509} & \numprint{2.30} & \textbf{\numprint{235875181}} & \textcolor{darkgreen}{\numprint{432.26}} & \numprint{3.36} & \textbf{\numprint{235875181}} \\ 
\hline 
\rule{0pt}{3ex}Solved instances & \multicolumn{2}{r|}{} & \multicolumn{3}{r|}{} & \multicolumn{2}{r|}{} & \multicolumn{3}{r|}{} & \multicolumn{2}{r}{80.65\% (25/31)  } \\ 
Optimal weight & \multicolumn{2}{r|}{28.00\% (7/25)  }  & \multicolumn{3}{r|}{84.00\% (21/25)  }  & \multicolumn{2}{r|}{68.00\% (17/25)  }  & \multicolumn{3}{r|}{88.00\% (22/25)  } & \multicolumn{2}{r}{} \\ 
\end{tabular} 

\caption{Best solution found by each algorithm and time (in seconds) required to compute it. $S_\text{base} = \frac{t_\text{base}}{t_\text{modified}}$ denotes the speedup between the modified and base versions of each local search. The global best solution is highlighted in \textbf{bold}. Rows are highlighted in gray if B~\&~R is able to find an exact solution.
}
\label{tab:reduce_rt}
\vspace*{-.25cm}
\end{table*}

For the SNAP networks, we see that B~\&~R solves $25$ of the $31$ instances we tested~\footnote{Using a longer time limit of $48$ hours we are able to solve 27 our of 31 instances.}.
Most notable, on seven of these instances where either HILS or DynWVC1 also find a solution with optimal weight, it does so up to two orders of magnitude faster.
This difference in performance compared to the OSM networks can be explained by the significantly lower graph density and less uniform degree distribution of the SNAP networks.
  These structural differences seem to allow for our reduction rules to be applicable more often, resulting in a significantly smaller kernel (as seen in Table~\ref{tab:props}). This is similar to the behavior of unweighted branch-and-reduce~\cite{akiba-tcs-2016}.
Therefore, except for a single instance, our algorithm is able to find the best solution on \emph{all} graphs tested.

Comparing the heuristic approaches, both versions of DynWVC perform better than HILS on most instances, with DynWVC2 often finding better solution than DynWVC1.
Nonetheless, HILS finds higher weight solutions than DynWVC1 and DynWVC2.

\subsection{The Power of Weighted Reductions.}
\label{sec:sota_reductions}
We now examine the effect of using reductions to improve existing heuristic algorithms.
For this purpose, we compare the combined approaches Red + HILS and Red + DynWVC with their base versions as well as our branch-and-reduce algorithm.
Our sample of results for the OSM and SNAP networks is given in Table~\ref{tab:reduce_rt}.
In addition to the data used in our state-of-the-art comparison, we now also report speedups between the modified and base versions of each local search.
Additionally, we give the percentage of instances solved by B~\&~R, as well as the percentage of solutions with optimal weight found be the inexact algorithms compared to B~\&~R.
For a full overview of all instances, we refer to Table~\ref{tab:osm_reduce_rt} and Table~\ref{tab:snap_reduce_rt} respectively.

When looking at the speedups for the SNAP graphs, we can see that using reductions allows local search to find optimal solutions orders of magnitude faster.
Additionally, they are now able to find an optimal solution more often than without reductions.
DynWVC2 in particular achieves an increase of $56\%$ of optimal solutions when using reductions.
Overall, we achieve a speedup of up to three orders of magnitude for the SNAP instances.
Thus, the additional costs for computing the kernel can be neglected for these instances.
However, for the OSM instances our reduction rules are less applicable and reducing the kernel comes at a significant cost compared to the unmodified local searches.

To further examine the influence of using reductions, Figure~\ref{fig:convergence} shows the solution quality over time for all algorithms and four instances.
For additional convergence plots, we refer to Figure~\ref{fig:apdxconvergence}.
For the OSM instances, we can see that initially DynWVC and HILS are able find good quality solutions much faster compared to their combined approaches.
However, once the kernel has been computed, regular DynWVC and HILS are quickly outperformed by the hybrid algorithms.

A more drastic change can be seen for the SNAP instances.
Instances were both DynWVC and HILS examine poor performance, Red~+~DynWVC and Red~+~HILS now rival our branch-and-reduce algorithm and give near-optimal solutions in less time.
Thus, using reductions for instances that are too large for traditional heuristic approaches allows for a~drastic~improvement.

\section{Conclusion and Future Work}
\label{sec:conclusion}
In this paper, we engineered a new branch-and-reduce algorithm as well as a combination of kernelization with local search for the maximum weight independent set problem. 
The core of our algorithms are a full suite of new reductions for the maximum weight independent set problem.
We performed extensive experiments to show the effectiveness of our algorithms in practice on real-world graphs of up to millions of vertices and edges. 
Our experimental evaluation shows that our branch-and-reduce algorithm can solve many large real-world instances quickly in practice, and that kernelization has a highly positive effect on local search~algorithms. 

As HILS often finds optimal solutions in practice, important future works include using this algorithm for the lower bound computation within our branch-and-reduce algorithm.
Furthermore, we would like to extend our discussion on the effectiveness of novel reduction rules.
In particular, we want to evaluate how much quality we gain from applying each individual rule and how the order we apply them in changes the resulting kernel size.

\renewcommand\bibsection{\section*{\refname}}
\bibliographystyle{abbrvnat}
\bibliography{mwis}

\begin{thebibliography}{34}
\providecommand{\natexlab}[1]{#1}
\providecommand{\url}[1]{\texttt{#1}}
\expandafter\ifx\csname urlstyle\endcsname\relax
  \providecommand{\doi}[1]{doi: #1}\else
  \providecommand{\doi}{doi: \begingroup \urlstyle{rm}\Url}\fi

\bibitem[OSM()]{OSMWEB}
\emph{OpenStreetMap}.
\newblock URL \url{https://www.openstreetmap.org}.

\bibitem[Ageev(1994)]{ageev1994finding}
A.~A. Ageev.
\newblock On finding critical independent and vertex sets.
\newblock \emph{SIAM Journal on Discrete Mathematics}, 7\penalty0 (2):\penalty0
  293--295, 1994.

\bibitem[Akiba and Iwata(2016)]{akiba-tcs-2016}
T.~Akiba and Y.~Iwata.
\newblock Branch-and-reduce exponential/{FPT} algorithms in practice: A case
  study of vertex cover.
\newblock \emph{Theoretical Computer Science}, 609, Part 1:\penalty0 211--225,
  2016.
\newblock \doi{10.1016/j.tcs.2015.09.023}.

\bibitem[Andrade et~al.(2012)Andrade, Resende, and Werneck]{andrade-2012}
D.~V. Andrade, M.~G. Resende, and R.~F. Werneck.
\newblock Fast local search for the maximum independent set problem.
\newblock \emph{Journal of Heuristics}, 18\penalty0 (4):\penalty0 525--547,
  2012.
\newblock \doi{10.1007/s10732-012-9196-4}.

\bibitem[Babel(1994)]{babel1994fast}
L.~Babel.
\newblock A fast algorithm for the maximum weight clique problem.
\newblock \emph{Computing}, 52\penalty0 (1):\penalty0 31--38, 1994.

\bibitem[Balas and Yu(1986)]{balas1986finding}
E.~Balas and C.~S. Yu.
\newblock Finding a maximum clique in an arbitrary graph.
\newblock \emph{SIAM Journal on Computing}, 15\penalty0 (4):\penalty0
  1054--1068, 1986.

\bibitem[Barth et~al.(2016)Barth, Niedermann, N\"{o}llenburg, and
  Strash]{barth-2016}
L.~Barth, B.~Niedermann, M.~N\"{o}llenburg, and D.~Strash.
\newblock Temporal map labeling: A new unified framework with experiments.
\newblock In \emph{Proceedings of the 24th ACM SIGSPATIAL International
  Conference on Advances in Geographic Information Systems}, GIS '16, pages
  23:1--23:10. ACM, 2016.
\newblock \doi{10.1145/2996913.2996957}.

\bibitem[Benlic and Hao(2013)]{benlic2013breakout}
U.~Benlic and J.-K. Hao.
\newblock Breakout local search for the quadratic assignment problem.
\newblock \emph{Applied Mathematics and Computation}, 219\penalty0
  (9):\penalty0 4800--4815, 2013.

\bibitem[Br{\'{e}}laz(1979)]{brelazcoloring}
D.~Br{\'{e}}laz.
\newblock New methods to color vertices of a graph.
\newblock \emph{Commun. {ACM}}, 22\penalty0 (4):\penalty0 251--256, 1979.
\newblock \doi{10.1145/359094.359101}.

\bibitem[Butenko and Trukhanov(2007)]{butenko-trukhanov}
S.~Butenko and S.~Trukhanov.
\newblock Using critical sets to solve the maximum independent set problem.
\newblock \emph{Operations Research Letters}, 35\penalty0 (4):\penalty0
  519--524, 2007.
\newblock \doi{10.1016/j.orl.2006.07.004}.

\bibitem[Butenko et~al.(2009)Butenko, Pardalos, Sergienko, Shylo, and
  Stetsyuk]{butenko-correcting-codes-2009}
S.~Butenko, P.~Pardalos, I.~Sergienko, V.~Shylo, and P.~Stetsyuk.
\newblock Estimating the size of correcting codes using extremal graph
  problems.
\newblock In C.~Pearce and E.~Hunt, editors, \emph{Optimization}, volume~32 of
  \emph{Springer Optimization and Its Applications}, pages 227--243. Springer,
  2009.
\newblock \doi{10.1007/978-0-387-98096-6_12}.

\bibitem[Cai and Lin(2016)]{cai2016fast}
S.~Cai and J.~Lin.
\newblock Fast solving maximum weight clique problem in massive graphs.
\newblock In \emph{Proceedings of the Twenty-Fifth International Joint
  Conference on Artificial Intelligence}, pages 568--574. AAAI Press, 2016.
\newblock URL \url{http://www.ijcai.org/Proceedings/16/Papers/087.pdf}.

\bibitem[Cai et~al.(2018)Cai, Hou, Lin, and Li]{cai-dynwvc}
S.~Cai, W.~Hou, J.~Lin, and Y.~Li.
\newblock Improving local search for minimum weight vertex cover by dynamic
  strategies.
\newblock In \emph{Proceedings of the Twenty-Seventh International Joint
  Conference on Artificial Intelligence ({IJCAI} 2018)}, pages 1412--1418,
  2018.
\newblock \doi{10.24963/ijcai.2018/196}.

\bibitem[Chang et~al.(2017)Chang, Li, and Zhang]{chang2017}
L.~Chang, W.~Li, and W.~Zhang.
\newblock Computing a near-maximum independent set in linear time by
  reducing-peeling.
\newblock \emph{Proceedings of the 2017 ACM International Conference on
  Management of Data (SIGMOD '17)}, pages 1181--1196, 2017.
\newblock \doi{10.1145/3035918.3035939}.

\bibitem[Chen et~al.(2001)Chen, Kanj, and Jia]{chen1999}
J.~Chen, I.~A. Kanj, and W.~Jia.
\newblock Vertex cover: Further observations and further improvements.
\newblock \emph{Journal of Algorithms}, 41\penalty0 (2):\penalty0 280--301,
  2001.
\newblock \doi{10.1006/jagm.2001.1186}.

\bibitem[Dahlum et~al.(2016)Dahlum, Lamm, Sanders, Schulz, Strash, and
  Werneck]{dahlum2016}
J.~Dahlum, S.~Lamm, P.~Sanders, C.~Schulz, D.~Strash, and R.~F. Werneck.
\newblock Accelerating local search for the maximum independent set problem.
\newblock In A.~V. Goldberg and A.~S. Kulikov, editors, \emph{Experimental
  Algorithms (SEA 2016)}, volume 9685 of \emph{LNCS}, pages 118--133. Springer,
  2016.
\newblock \doi{10.1007/978-3-319-38851-9_9}.

\bibitem[Fomin et~al.(2009)Fomin, Grandoni, and Kratsch]{fomin-2009}
F.~V. Fomin, F.~Grandoni, and D.~Kratsch.
\newblock A measure \& conquer approach for the analysis of exact algorithms.
\newblock \emph{J. ACM}, 56\penalty0 (5):\penalty0 25:1--25:32, 2009.
\newblock \doi{10.1145/1552285.1552286}.

\bibitem[Garey et~al.(1974)Garey, Johnson, and Stockmeyer]{garey1974}
M.~R. Garey, D.~S. Johnson, and L.~Stockmeyer.
\newblock {Some Simplified {N}{P}-Complete Problems}.
\newblock In \emph{Proceedings of the 6th ACM Symposium on Theory of
  Computing}, STOC '74, pages 47--63. ACM, 1974.

\bibitem[Gemsa et~al.(2014)Gemsa, N\"ollenburg, and
  Rutter]{gemsa2014dynamiclabel}
A.~Gemsa, M.~N\"ollenburg, and I.~Rutter.
\newblock Evaluation of labeling strategies for rotating maps.
\newblock In \emph{Experimental Algorithms (SEA'14)}, volume 8504 of
  \emph{LNCS}, pages 235--246. Springer, 2014.
\newblock \doi{10.1007/978-3-319-07959-2_20}.

\bibitem[Hespe et~al.(2018)Hespe, Schulz, and Strash]{hespe2018scalable}
D.~Hespe, C.~Schulz, and D.~Strash.
\newblock Scalable kernelization for maximum independent sets.
\newblock In \emph{2018 Proceedings of the Twentieth Workshop on Algorithm
  Engineering and Experiments (ALENEX)}, pages 223--237. SIAM, 2018.
\newblock \doi{10.1137/1.9781611975055.19}.

\bibitem[Iwata et~al.(2014)Iwata, Oka, and Yoshida]{iwata-2014}
Y.~Iwata, K.~Oka, and Y.~Yoshida.
\newblock Linear-time {FPT} algorithms via network flow.
\newblock In \emph{Proc. 25th ACM-SIAM Symposium on Discrete Algorithms}, SODA
  '14, pages 1749--1761. SIAM, 2014.
\newblock \doi{10.1137/1.9781611973402.127}.

\bibitem[Lamm et~al.(2017)Lamm, Sanders, Schulz, Strash, and
  Werneck]{redumis-2017}
S.~Lamm, P.~Sanders, C.~Schulz, D.~Strash, and R.~F. Werneck.
\newblock Finding near-optimal independent sets at scale.
\newblock \emph{Journal of Heuristics}, 23\penalty0 (4):\penalty0 207--229,
  2017.
\newblock \doi{10.1007/s10732-017-9337-x}.

\bibitem[Larson(2007)]{larson-2007}
C.~Larson.
\newblock A note on critical independence reductions.
\newblock volume~51 of \emph{Bulletin of the Institute of Combinatorics and its
  Applications}, pages 34--46, 2007.

\bibitem[Leskovec and Krevl(2014)]{snapnets}
J.~Leskovec and A.~Krevl.
\newblock {SNAP Datasets}: {Stanford} large network dataset collection.
\newblock \url{http://snap.stanford.edu/data}, June 2014.

\bibitem[Li et~al.(2017)Li, Cai, and Hou]{li2017efficient}
Y.~Li, S.~Cai, and W.~Hou.
\newblock An efficient local search algorithm for minimum weighted vertex cover
  on massive graphs.
\newblock In \emph{Asia-Pacific Conference on Simulated Evolution and Learning
  (SEAL 2017)}, volume 10593 of \emph{LNCS}, pages 145--157. 2017.
\newblock \doi{10.1007/978-3-319-68759-9_13}.

\bibitem[Nemhauser and Trotter(1975)]{nemhauser-1975}
G.~Nemhauser and J.~Trotter, L.E.
\newblock Vertex packings: Structural properties and algorithms.
\newblock \emph{Mathematical Programming}, 8\penalty0 (1):\penalty0 232--248,
  1975.
\newblock \doi{10.1007/BF01580444}.

\bibitem[Nogueira et~al.(2018)Nogueira, Pinheiro, and
  Subramanian]{hybrid-ils-2018}
B.~Nogueira, R.~G.~S. Pinheiro, and A.~Subramanian.
\newblock A hybrid iterated local search heuristic for the maximum weight
  independent set problem.
\newblock \emph{Optimization Letters}, 12\penalty0 (3):\penalty0 567--583,
  2018.
\newblock \doi{10.1007/s11590-017-1128-7}.

\bibitem[{\"O}sterg{\aa}rd(2002)]{ostergaard2002fast}
P.~R. {\"O}sterg{\aa}rd.
\newblock A fast algorithm for the maximum clique problem.
\newblock \emph{Discrete Applied Mathematics}, 120\penalty0 (1-3):\penalty0
  197--207, 2002.

\bibitem[Pullan(2006)]{pullan-2006}
W.~Pullan.
\newblock Phased local search for the maximum clique problem.
\newblock \emph{J. Comb. Optim.}, 12\penalty0 (3):\penalty0 303--323, 2006.
\newblock \doi{10.1007/s10878-006-9635-y}.

\bibitem[Pullan(2009)]{pullan-2009}
W.~Pullan.
\newblock Optimisation of unweighted/weighted maximum independent sets and
  minimum vertex covers.
\newblock \emph{Discrete Optim.}, 6\penalty0 (2):\penalty0 214--219, 2009.
\newblock ISSN 1572-5286.
\newblock \doi{10.1016/j.disopt.2008.12.001}.

\bibitem[Strash(2016)]{strash-power-2016}
D.~Strash.
\newblock On the power of simple reductions for the maximum independent set
  problem.
\newblock In T.~N. Dinh and M.~T. Thai, editors, \emph{Computing and
  Combinatorics (COCOON'16)}, volume 9797 of \emph{LNCS}, pages 345--356. 2016.
\newblock \doi{10.1007/978-3-319-42634-1_28}.

\bibitem[Warren and Hicks(2006)]{warren2006combinatorial}
J.~S. Warren and I.~V. Hicks.
\newblock Combinatorial branch-and-bound for the maximum weight independent set
  problem.
\newblock 2006.
\newblock URL \url{https://www.caam.rice.edu/~ivhicks/jeff.rev.pdf}.

\bibitem[Wu et~al.(2012)Wu, Hao, and Glover]{wu2012multi}
Q.~Wu, J.-K. Hao, and F.~Glover.
\newblock Multi-neighborhood tabu search for the maximum weight clique problem.
\newblock \emph{Annals of Operations Research}, 196\penalty0 (1):\penalty0
  611--634, 2012.

\bibitem[Xu et~al.(2016)Xu, Kumar, and Koenig]{xu2016new}
H.~Xu, T.~S. Kumar, and S.~Koenig.
\newblock A new solver for the minimum weighted vertex cover problem.
\newblock In \emph{International Conference on AI and OR Techniques in
  Constriant Programming for Combinatorial Optimization Problems}, pages
  392--405. Springer, 2016.

\end{thebibliography}

\begin{appendix}
\section{Omitted Proofs}
\setcounter{reduction}{2}
\begin{reduction}[Neighborhood Folding]
Let $v \in V$, and suppose that $N(v)$ is independent. If $w(N(v)) > w(v)$, but $w(N(v)) - \min_{u\in N(v)}\{w(u)\} < w(v)$, then fold $v$ and $N(v)$ into a new vertex $v'$ with weight $w(v') = w(N(v)) - w(v)$. Let $\I'$ be an MWIS of $G'$, then we construct an MWIS $\I$ of $G$ as follows: If $v'\in \I'$ then $\I = (\I'\setminus\{v'\}) \cup N(v)$, otherwise if $v\in \I'$ then $\I = \I' \cup \{v\}$. Furthermore, $\alpha_w(G) = \alpha_w(G') + w(v)$.
\end{reduction}
\begin{proof}
\label{ommittedproofs}
First note that after folding, the following graphs are identical: $G'[V'\setminus N_{G'}[v']] = G[V\setminus N[N[v]]$ and $G'[V'\setminus \{v'\}] = G[V\setminus N[v]]$.
Let $\I'$ be an MWIS of $G'$. We have two cases.

\noindent \emph{Case 1 ($v'\in\I'$):}
Suppose that $v'\in\I'$. We show that $w(N(v)) + \alpha_w(G[V\setminus N[N[v]]]) \geq w(v) + \alpha_w(G[V\setminus N[v]])$, which shows that the vertices of $N(v)$ are together in some MWIS of $G$.
 Since $v'\in\I'$, we have that
\begin{align*}
w(v) + \alpha_w(G') &= w(v) + w(v') + \alpha_w(G'[V'\setminus N_{G'}[v']])\\
                    &= w(v) + w(N(v)) - w(v) \\
                    &\phantom{= w(v) + w(N(v))\text{ }} + \alpha_w(G'[V'\setminus N_{G'}[v']])\\
                    &= w(N(v)) + \alpha_w(G[V\setminus N[N[v]]]).
\end{align*}
%

\noindent But since $\I'$ is an MWIS of $G'$, we have that
\begin{align*}
w(v) + \alpha_w(G') &\geq w(v) + \alpha_w(G'[V'\setminus \{v'\}]) \\
                    &= w(v) + \alpha_w(G[V\setminus N[v]]).
\end{align*}

\noindent Thus,
$w(N(v)) + \alpha_w(G[V\setminus N[N[v]]]) \geq w(v) + \alpha_w(G[V\setminus N[v]])$ and
the vertices of $N(v)$ are together in some MWIS of $G$.
Furthermore, 
we have that
\begin{align*}
\alpha_w(G) &= w(N(v)) + \alpha_w(G[V\setminus N[N[v]]]) \\ &= \alpha_w(G') + w(v).
\end{align*}

\noindent \emph{Case 2: ($v'\notin\I'$):}
Suppose that $v'\notin\I'$. We show that $w(v) + \alpha_w(G[V\setminus N[v]]) \geq w(N(v)) + \alpha_w(G[V\setminus N[N[v]]])$, which shows that $v$ is in some MWIS of $G$.
Since $v'\notin\I'$, we have that
\begin{align*}
w(v) + \alpha_w(G') &= w(v) + \alpha_w(G'[V'\setminus \{v'\}])\\
                    &= w(v) + \alpha_w(G[V\setminus N[v]])
\end{align*}

\noindent But since $\I'$ is an MWIS of $G'$, we have that
\begin{align*}
w(v) + \alpha_w(G') &\geq w(v) + w(v') + \alpha_w(G'[V'\setminus N_{G'}[v']]) \\
                    &= w(v) + w(N(v)) - w(v) \\
                    &\phantom{= w(v) + w(N(v))\text{ } }+ \alpha_w(G[V\setminus N[N[v]]]) \\
                    &= w(N(v)) + \alpha_w(G[V\setminus N[N[v]]]).
\end{align*}

\noindent Thus,
$w(v) + \alpha_w(G[V\setminus N[v]]) \geq w(u) + w(x) + \alpha_w(G[V\setminus N[\{u,x\}]])$ and
$v$ is in some MWIS of $G$.
Lastly,
\begin{align*}
\alpha_w(G) &= w(v) + \alpha_w(G[V\setminus N[v]]) \\ &= \alpha_w(G') + w(v).
\end{align*}
\end{proof}

\setcounter{reduction}{5}
\begin{reduction}[Isolated weight transfer]
Let $v\in V$ be isolated, and suppose that the set of isolated vertices $S(v)\subseteq N(v)$ is such that $\forall u\in S(v)$, $w(v) \geq w(u)$. We
\begin{enumerate}[(i)]
\item remove all $u\in N(v)$ such that $w(u)\leq w(v)$, and let the remaining neighbors be denoted by $N'(v)$,
\item remove $v$ and $\forall x\in N'(v)$ set its new weight to $w'(x) = w(x) - w(v)$, and
\end{enumerate}
let the resulting graph be denoted by $G'$. Then $\alpha_w(G) = w(v) + \alpha_w(G')$ and an MWIS $\I$ of $G$ can be constructed from an MWIS $\I'$ of $G'$ as follows: if $\I' \cap N'(v) = \emptyset$ then $\I = \I'\cup\{v\}$, otherwise $\I = \I'$.
\end{reduction}
\begin{proof}
For (i), note that it is safe to remove all $u\in N(v)$ such that $w(u)\leq w(v)$ since these vertices meet the criteria for the neighbor removal reduction. All vertices $x\in N'(v)$ that remain have weight $w(x) > w(v)$ and are not isolated.

\emph{Case 1 ($\I'\cap N'(v) = \emptyset$):} 
Let $\I'$ be an MWIS of $G'$, we show that if $\I'\cap N'(v) = \emptyset$ then $I = I'\cup\{v\}$. 
To show this, we show that $w(v) + \alpha_w(G[V\setminus N[v]]) \geq \alpha_w(G[V\setminus\{v\}])$.

Let $x\in N'(v)$. Since $x\notin I'$, we have that 
\begin{align*}
w(v) + \alpha_w(G') &= w(v) + \alpha_w(G'[V'\setminus N'(v)]) \\
                    &= w(v) + \alpha_w(G[V\setminus N[v]])
\end{align*}
and
\begin{align*}
w(v) + \alpha_w(G') &\geq  w(v) + w'(x) + \alpha_w(G'[V'\setminus N[x]]) \\
                    &= w(v) + w(x) - w(v) + \alpha_w(G'[V'\setminus N[x]])\\
                    &= w(x) + \alpha_w(G[V\setminus N[x]]).
\end{align*}

Thus, for any $x\in N'(v)$, we have that
\begin{align*}
w(v) + \alpha_w(G') &= w(v) + \alpha_w(G[V\setminus N[v]]) \\
                    &\geq w(x) + \alpha_w(G[V\setminus N[x]])
\end{align*}
and therefore the heaviest independent set containing $v$ is at least the weight of the heaviest independent containing \emph{any} neighbor of $v$. Concluding, we have that
\[w(v) + \alpha_w(G[V\setminus N[v]]) \geq \alpha_w(G[V\setminus\{v\}])\]
and therefore $\I = \I'\cup\{v\}$ is an MWIS of $G$.

\emph{Case 2 ($\I'\cap N'(v) \neq \emptyset$):} 
Let $\I'$ be an MWIS of $G'$, we show that if $\I' \cap N'(v) \neq \emptyset$ then $\I = \I'$.
To show this, let $\{x\} = \I'\cap N'(v)$.
Define $G''$ as the graph resulting from increasing the weight of $N'(v)$ by $w(v)$, \ie~$\forall~u~\in~N'(v)$ we set $w''(u) = w'(u) + w(v) = w(u)$.
We first show that $\I'' = \I'$ is an MWIS of $G'$.
Therefore, assume that $\I^*$ is an MWIS of $G''$ with $w(\I^*) > w(\I')$ that does not contain $x$.
However, then $\I^*$ is also a better MWIS on $G'$ which contradicts our initial assumption.
Finally, we have that $w(\I'') = w(\I') + w(v)$, since exactly one node in $N'(v)$ is in $\I'$.

Next, we define $G'''$  as the graph resulting from adding back $v$ to $G'$' and show that $\I''' = \I''$ is a MWIS of $G'''$.
For this purpose, we assume that $\I^*$ is a MWIS of $G'''$ with $w(\I^*) > w(\I''')$.
Then, $v \in \I^*$ since we only added this node to $G''$.
Likewise, $x \not\in \I^*$ since its a neighbor of $v$.

Since $w(\I^*) > w(\I''')$, we have that:
\begin{align*}
  w(\I^*\setminus\{v\})  &= w(\I^*) - w(v)\\
                &> w(\I'') - w(v)\\
                &= w(\I') + w(v) - w(v)\\
                &= w(\I').
\end{align*}
However, since $\I^*\setminus\{v\}$ does neither include v nor any neighbor of $v$ it is also an IS of $G'$ that is larger than $\I'$.
This contradicts our initial assumption and thus $\I''' = \I'' = \I'$.
Furthermore, since $G''' = G$, we have that $\I''' = \I = \I'$.
\end{proof}

\vfill\pagebreak

\section{Graph Properties, Kernel Sizes, Full Tables, Convergence Plots}
\begin{table}[h!]
\scriptsize
\centering
\vspace*{-.5cm}
\begin{tabular}{l|r r r r} 
Graph & $|V|$ & $|E|$ & $\mathcal{K}_{\text{dense}}$ & $\mathcal{K}_{\text{full}}$ \\ 
\hline 
\Id{\detokenize{alabama-AM2}} & \numprint{1164} & \numprint{38772} & \numprint{173} & \numprint{173} \\ 
\Id{\detokenize{alabama-AM3}} & \numprint{3504} & \numprint{619328} & \numprint{1614} & \numprint{1614} \\ 
\Id{\detokenize{district-of-columbia-AM1}} & \numprint{2500} & \numprint{49302} & \numprint{800} & \numprint{800} \\ 
\Id{\detokenize{district-of-columbia-AM2}} & \numprint{13597} & \numprint{3219590} & \numprint{6360} & \numprint{6360} \\ 
\Id{\detokenize{district-of-columbia-AM3}} & \numprint{46221} & \numprint{55458274} & \numprint{33367} & \numprint{33367} \\ 
\Id{\detokenize{florida-AM2}} & \numprint{1254} & \numprint{33872} & \numprint{41} & \numprint{41} \\ 
\Id{\detokenize{florida-AM3}} & \numprint{2985} & \numprint{308086} & \numprint{1069} & \numprint{1069} \\ 
\Id{\detokenize{georgia-AM3}} & \numprint{1680} & \numprint{148252} & \numprint{861} & \numprint{861} \\ 
\Id{\detokenize{greenland-AM3}} & \numprint{4986} & \numprint{7304722} & \numprint{3942} & \numprint{3942} \\ 
\Id{\detokenize{hawaii-AM2}} & \numprint{2875} & \numprint{530316} & \numprint{428} & \numprint{428} \\ 
\Id{\detokenize{hawaii-AM3}} & \numprint{28006} & \numprint{98889842} & \numprint{24436} & \numprint{24436} \\ 
\Id{\detokenize{idaho-AM3}} & \numprint{4064} & \numprint{7848160} & \numprint{3208} & \numprint{3208} \\ 
\Id{\detokenize{kansas-AM3}} & \numprint{2732} & \numprint{1613824} & \numprint{1605} & \numprint{1605} \\ 
\Id{\detokenize{kentucky-AM2}} & \numprint{2453} & \numprint{1286856} & \numprint{442} & \numprint{442} \\ 
\Id{\detokenize{kentucky-AM3}} & \numprint{19095} & \numprint{119067260} & \numprint{16871} & \numprint{16871} \\ 
\Id{\detokenize{louisiana-AM3}} & \numprint{1162} & \numprint{74154} & \numprint{382} & \numprint{382} \\ 
\Id{\detokenize{maryland-AM3}} & \numprint{1018} & \numprint{190830} & \numprint{187} & \numprint{187} \\ 
\Id{\detokenize{massachusetts-AM2}} & \numprint{1339} & \numprint{70898} & \numprint{196} & \numprint{196} \\ 
\Id{\detokenize{massachusetts-AM3}} & \numprint{3703} & \numprint{1102982} & \numprint{2008} & \numprint{2008} \\ 
\Id{\detokenize{mexico-AM3}} & \numprint{1096} & \numprint{94262} & \numprint{620} & \numprint{620} \\ 
\Id{\detokenize{new-hampshire-AM3}} & \numprint{1107} & \numprint{36042} & \numprint{247} & \numprint{247} \\ 
\Id{\detokenize{north-carolina-AM3}} & \numprint{1557} & \numprint{473478} & \numprint{1178} & \numprint{1178} \\ 
\Id{\detokenize{oregon-AM2}} & \numprint{1325} & \numprint{115034} & \numprint{35} & \numprint{35} \\ 
\Id{\detokenize{oregon-AM3}} & \numprint{5588} & \numprint{5825402} & \numprint{3670} & \numprint{3670} \\ 
\Id{\detokenize{pennsylvania-AM3}} & \numprint{1148} & \numprint{52928} & \numprint{315} & \numprint{315} \\ 
\Id{\detokenize{rhode-island-AM2}} & \numprint{2866} & \numprint{590976} & \numprint{1103} & \numprint{1103} \\ 
\Id{\detokenize{rhode-island-AM3}} & \numprint{15124} & \numprint{25244438} & \numprint{13031} & \numprint{13031} \\ 
\Id{\detokenize{utah-AM3}} & \numprint{1339} & \numprint{85744} & \numprint{568} & \numprint{568} \\ 
\Id{\detokenize{vermont-AM3}} & \numprint{3436} & \numprint{2272328} & \numprint{2630} & \numprint{2630} \\ 
\Id{\detokenize{virginia-AM2}} & \numprint{2279} & \numprint{120080} & \numprint{237} & \numprint{237} \\ 
\Id{\detokenize{virginia-AM3}} & \numprint{6185} & \numprint{1331806} & \numprint{3867} & \numprint{3867} \\ 
\Id{\detokenize{washington-AM2}} & \numprint{3025} & \numprint{304898} & \numprint{382} & \numprint{382} \\ 
\Id{\detokenize{washington-AM3}} & \numprint{10022} & \numprint{4692426} & \numprint{8030} & \numprint{8030} \\ 
\Id{\detokenize{west-virginia-AM3}} & \numprint{1185} & \numprint{251240} & \numprint{991} & \numprint{991} \\ 
\end{tabular}

\vspace*{.5cm}
\begin{tabular}{l|r r r r} 
Graph & $|V|$ & $|E|$ & $\mathcal{K}_{\text{dense}}$ & $\mathcal{K}_{\text{full}}$ \\ 
\hline 
\Id{\detokenize{as-skitter}} & \numprint{1696415} & \numprint{22190596} & \numprint{27318} & \numprint{9180} \\ 
\Id{\detokenize{ca-AstroPh}} & \numprint{18772} & \numprint{396100} & \numprint{0} & \numprint{0} \\ 
\Id{\detokenize{ca-CondMat}} & \numprint{23133} & \numprint{186878} & \numprint{0} & \numprint{0} \\ 
\Id{\detokenize{ca-GrQc}} & \numprint{5242} & \numprint{28968} & \numprint{0} & \numprint{0} \\ 
\Id{\detokenize{ca-HepPh}} & \numprint{12008} & \numprint{236978} & \numprint{0} & \numprint{0} \\ 
\Id{\detokenize{ca-HepTh}} & \numprint{9877} & \numprint{51946} & \numprint{0} & \numprint{0} \\ 
\Id{\detokenize{email-Enron}} & \numprint{36692} & \numprint{367662} & \numprint{0} & \numprint{0} \\ 
\Id{\detokenize{email-EuAll}} & \numprint{265214} & \numprint{728962} & \numprint{0} & \numprint{0} \\ 
\Id{\detokenize{p2p-Gnutella04}} & \numprint{10876} & \numprint{79988} & \numprint{0} & \numprint{0} \\ 
\Id{\detokenize{p2p-Gnutella05}} & \numprint{8846} & \numprint{63678} & \numprint{0} & \numprint{0} \\ 
\Id{\detokenize{p2p-Gnutella06}} & \numprint{8717} & \numprint{63050} & \numprint{0} & \numprint{0} \\ 
\Id{\detokenize{p2p-Gnutella08}} & \numprint{6301} & \numprint{41554} & \numprint{0} & \numprint{0} \\ 
\Id{\detokenize{p2p-Gnutella09}} & \numprint{8114} & \numprint{52026} & \numprint{0} & \numprint{0} \\ 
\Id{\detokenize{p2p-Gnutella24}} & \numprint{26518} & \numprint{130738} & \numprint{0} & \numprint{0} \\ 
\Id{\detokenize{p2p-Gnutella25}} & \numprint{22687} & \numprint{109410} & \numprint{11} & \numprint{0} \\ 
\Id{\detokenize{p2p-Gnutella30}} & \numprint{36682} & \numprint{176656} & \numprint{10} & \numprint{0} \\ 
\Id{\detokenize{p2p-Gnutella31}} & \numprint{62586} & \numprint{295784} & \numprint{0} & \numprint{0} \\ 
\Id{\detokenize{roadNet-CA}} & \numprint{1965206} & \numprint{5533214} & \numprint{233083} & \numprint{63926} \\ 
\Id{\detokenize{roadNet-PA}} & \numprint{1088092} & \numprint{3083796} & \numprint{135536} & \numprint{38080} \\ 
\Id{\detokenize{roadNet-TX}} & \numprint{1379917} & \numprint{3843320} & \numprint{151570} & \numprint{39433} \\ 
\Id{\detokenize{soc-Epinions1}} & \numprint{75879} & \numprint{811480} & \numprint{6} & \numprint{0} \\ 
\Id{\detokenize{soc-LiveJournal1}} & \numprint{4847571} & \numprint{85702474} & \numprint{61690} & \numprint{29779} \\ 
\Id{\detokenize{soc-Slashdot0811}} & \numprint{77360} & \numprint{938360} & \numprint{0} & \numprint{0} \\ 
\Id{\detokenize{soc-Slashdot0902}} & \numprint{82168} & \numprint{1008460} & \numprint{20} & \numprint{0} \\ 
\Id{\detokenize{soc-pokec-relationships}} & \numprint{1632803} & \numprint{44603928} & \numprint{927214} & \numprint{902748} \\ 
\Id{\detokenize{web-BerkStan}} & \numprint{685230} & \numprint{13298940} & \numprint{37004} & \numprint{17482} \\ 
\Id{\detokenize{web-Google}} & \numprint{875713} & \numprint{8644102} & \numprint{2892} & \numprint{1178} \\ 
\Id{\detokenize{web-NotreDame}} & \numprint{325729} & \numprint{2180216} & \numprint{14038} & \numprint{6760} \\ 
\Id{\detokenize{web-Stanford}} & \numprint{281903} & \numprint{3985272} & \numprint{14280} & \numprint{2640} \\ 
\Id{\detokenize{wiki-Talk}} & \numprint{2394385} & \numprint{9319130} & \numprint{0} & \numprint{0} \\ 
\Id{\detokenize{wiki-Vote}} & \numprint{7115} & \numprint{201524} & \numprint{246} & \numprint{237} \\ 
\end{tabular} 

\caption{Basic properties as well as kernel sizes computed by both variants of our branch-and-reduce algorithm for the OSM networks (top) and SNAP networks (bottom).
}
\label{tab:props}
\end{table}

\begin{table*}[ht]
\scriptsize
\centering
\setlength{\tabcolsep}{0.5ex}
\begin{tabular}{l|r r|r r|r r|r r|r r} 
& \multicolumn{2}{c|}{DynWVC1} & \multicolumn{2}{c|}{DynWVC2} & \multicolumn{2}{c|}{HILS} & \multicolumn{2}{c|}{\BRD} & \multicolumn{2}{c}{\BRF} \\ 
\hline 
Graph & $t_\text{max}$ & $w_\text{max}$ & $t_\text{max}$ & $w_\text{max}$ & $t_\text{max}$ & $w_\text{max}$ & $t_\text{max}$ & $w_\text{max}$ & $t_\text{max}$ & $w_\text{max}$ \\ 
\hline 
\rowcolor{lightergray} \Id{\detokenize{alabama-AM2}} & \numprint{0.62} & \numprint{174241} & \numprint{26.83} & \numprint{174297} & \numprint{0.04} & \textbf{\numprint{174309}} & \numprint{0.40} & \textbf{\numprint{174309}} & \numprint{0.79} & \textbf{\numprint{174309}} \\ 
\Id{\detokenize{alabama-AM3}} & \numprint{464.02} & \numprint{185527} & \numprint{887.55} & \numprint{185652} & \numprint{0.73} & \textbf{\numprint{185744}} & \numprint{15.79} & \numprint{185707} & \numprint{80.78} & \numprint{185707} \\ 
\Id{\detokenize{district-of-columbia-AM1}} & \numprint{12.64} & \textbf{\numprint{196475}} & \numprint{11.40} & \textbf{\numprint{196475}} & \numprint{0.26} & \textbf{\numprint{196475}} & \numprint{1.97} & \textbf{\numprint{196475}} & \numprint{4.13} & \textbf{\numprint{196475}} \\ 
\Id{\detokenize{district-of-columbia-AM2}} & \numprint{272.37} & \numprint{208942} & \numprint{596.62} & \numprint{208954} & \numprint{717.75} & \textbf{\numprint{209132}} & \numprint{20.03} & \numprint{147450} & \numprint{233.70} & \numprint{147450} \\ 
\Id{\detokenize{district-of-columbia-AM3}} & \numprint{949.96} & \numprint{224289} & \numprint{782.62} & \numprint{223780} & \numprint{989.68} & \textbf{\numprint{227598}} & \numprint{553.84} & \numprint{92784} & \numprint{918.07} & \numprint{92714} \\ 
\rowcolor{lightergray} \Id{\detokenize{florida-AM2}} & \numprint{1.14} & \textbf{\numprint{230595}} & \numprint{0.72} & \textbf{\numprint{230595}} & \numprint{0.04} & \textbf{\numprint{230595}} & \numprint{0.03} & \textbf{\numprint{230595}} & \numprint{0.02} & \textbf{\numprint{230595}} \\ 
\rowcolor{lightergray} \Id{\detokenize{florida-AM3}} & \numprint{553.56} & \numprint{237127} & \numprint{181.58} & \numprint{237081} & \numprint{2.76} & \textbf{\numprint{237333}} & \numprint{20.52} & \textbf{\numprint{237333}} & \numprint{324.38} & \numprint{226767} \\ 
\Id{\detokenize{georgia-AM3}} & \numprint{0.88} & \textbf{\numprint{222652}} & \numprint{1.29} & \textbf{\numprint{222652}} & \numprint{0.05} & \textbf{\numprint{222652}} & \numprint{4.88} & \numprint{214918} & \numprint{14.35} & \numprint{214918} \\ 
\Id{\detokenize{greenland-AM3}} & \numprint{73.16} & \textbf{\numprint{14011}} & \numprint{51.09} & \numprint{14008} & \numprint{1.72} & \textbf{\numprint{14011}} & \numprint{14.52} & \numprint{13152} & \numprint{47.25} & \numprint{13069} \\ 
\rowcolor{lightergray} \Id{\detokenize{hawaii-AM2}} & \numprint{4.85} & \numprint{125273} & \numprint{3.20} & \numprint{125276} & \numprint{0.33} & \textbf{\numprint{125284}} & \numprint{3.59} & \textbf{\numprint{125284}} & \numprint{10.89} & \textbf{\numprint{125284}} \\ 
\Id{\detokenize{hawaii-AM3}} & \numprint{898.64} & \numprint{140596} & \numprint{904.15} & \numprint{140486} & \numprint{332.32} & \textbf{\numprint{141035}} & \numprint{288.58} & \numprint{106251} & \numprint{1177.95} & \numprint{129812} \\ 
\Id{\detokenize{idaho-AM3}} & \numprint{76.55} & \textbf{\numprint{77145}} & \numprint{85.35} & \textbf{\numprint{77145}} & \numprint{1.49} & \textbf{\numprint{77145}} & \numprint{866.90} & \numprint{77010} & \numprint{61.26} & \numprint{76831} \\ 
\Id{\detokenize{kansas-AM3}} & \numprint{46.87} & \textbf{\numprint{87976}} & \numprint{44.26} & \textbf{\numprint{87976}} & \numprint{0.84} & \textbf{\numprint{87976}} & \numprint{11.35} & \numprint{87925} & \numprint{18.99} & \numprint{87925} \\ 
\rowcolor{lightergray} \Id{\detokenize{kentucky-AM2}} & \numprint{5.12} & \textbf{\numprint{97397}} & \numprint{7.39} & \textbf{\numprint{97397}} & \numprint{0.47} & \textbf{\numprint{97397}} & \numprint{11.35} & \textbf{\numprint{97397}} & \numprint{42.05} & \textbf{\numprint{97397}} \\ 
\Id{\detokenize{kentucky-AM3}} & \numprint{932.32} & \numprint{100463} & \numprint{722.69} & \numprint{100430} & \numprint{802.03} & \textbf{\numprint{100507}} & \numprint{172.30} & \numprint{91864} & \numprint{3346.94} & \numprint{96634} \\ 
\rowcolor{lightergray} \Id{\detokenize{louisiana-AM3}} & \numprint{0.32} & \numprint{60005} & \numprint{0.27} & \numprint{60002} & \numprint{0.03} & \textbf{\numprint{60024}} & \numprint{3.38} & \textbf{\numprint{60024}} & \numprint{20.17} & \textbf{\numprint{60024}} \\ 
\rowcolor{lightergray} \Id{\detokenize{maryland-AM3}} & \numprint{1.34} & \textbf{\numprint{45496}} & \numprint{0.87} & \textbf{\numprint{45496}} & \numprint{0.02} & \textbf{\numprint{45496}} & \numprint{3.34} & \textbf{\numprint{45496}} & \numprint{11.08} & \textbf{\numprint{45496}} \\ 
\rowcolor{lightergray} \Id{\detokenize{massachusetts-AM2}} & \numprint{0.37} & \textbf{\numprint{140095}} & \numprint{0.09} & \textbf{\numprint{140095}} & \numprint{0.02} & \textbf{\numprint{140095}} & \numprint{0.46} & \textbf{\numprint{140095}} & \numprint{0.48} & \textbf{\numprint{140095}} \\ 
\Id{\detokenize{massachusetts-AM3}} & \numprint{435.31} & \numprint{145863} & \numprint{154.61} & \numprint{145863} & \numprint{2.73} & \textbf{\numprint{145866}} & \numprint{12.87} & \numprint{145617} & \numprint{23.97} & \numprint{145631} \\ 
\rowcolor{lightergray} \Id{\detokenize{mexico-AM3}} & \numprint{0.14} & \textbf{\numprint{97663}} & \numprint{46.86} & \textbf{\numprint{97663}} & \numprint{0.04} & \textbf{\numprint{97663}} & \numprint{14.25} & \textbf{\numprint{97663}} & \numprint{289.14} & \textbf{\numprint{97663}} \\ 
\rowcolor{lightergray} \Id{\detokenize{new-hampshire-AM3}} & \numprint{0.22} & \textbf{\numprint{116060}} & \numprint{0.42} & \textbf{\numprint{116060}} & \numprint{0.03} & \textbf{\numprint{116060}} & \numprint{3.25} & \textbf{\numprint{116060}} & \numprint{8.75} & \textbf{\numprint{116060}} \\ 
\Id{\detokenize{north-carolina-AM3}} & \numprint{796.26} & \numprint{49716} & \numprint{285.91} & \textbf{\numprint{49720}} & \numprint{0.08} & \textbf{\numprint{49720}} & \numprint{10.45} & \numprint{49562} & \numprint{11.55} & \numprint{49562} \\ 
\rowcolor{lightergray} \Id{\detokenize{oregon-AM2}} & \numprint{0.22} & \textbf{\numprint{165047}} & \numprint{0.25} & \textbf{\numprint{165047}} & \numprint{0.04} & \textbf{\numprint{165047}} & \numprint{0.04} & \textbf{\numprint{165047}} & \numprint{0.09} & \textbf{\numprint{165047}} \\ 
\Id{\detokenize{oregon-AM3}} & \numprint{393.23} & \numprint{175046} & \numprint{126.97} & \numprint{175060} & \numprint{3.36} & \textbf{\numprint{175078}} & \numprint{351.99} & \numprint{174334} & \numprint{474.15} & \numprint{164941} \\ 
\rowcolor{lightergray} \Id{\detokenize{pennsylvania-AM3}} & \numprint{0.09} & \textbf{\numprint{143870}} & \numprint{0.15} & \textbf{\numprint{143870}} & \numprint{0.04} & \textbf{\numprint{143870}} & \numprint{9.98} & \textbf{\numprint{143870}} & \numprint{38.76} & \textbf{\numprint{143870}} \\ 
\Id{\detokenize{rhode-island-AM2}} & \numprint{6.66} & \numprint{184562} & \numprint{24.74} & \numprint{184576} & \numprint{0.40} & \textbf{\numprint{184596}} & \numprint{10.70} & \numprint{184543} & \numprint{16.79} & \numprint{184543} \\ 
\Id{\detokenize{rhode-island-AM3}} & \numprint{54.99} & \numprint{201553} & \numprint{609.14} & \numprint{201344} & \numprint{43.34} & \textbf{\numprint{201758}} & \numprint{399.33} & \numprint{162639} & \numprint{931.05} & \numprint{163080} \\ 
\rowcolor{lightergray} \Id{\detokenize{utah-AM3}} & \numprint{136.15} & \numprint{98802} & \numprint{233.52} & \textbf{\numprint{98847}} & \numprint{0.08} & \textbf{\numprint{98847}} & \numprint{64.04} & \textbf{\numprint{98847}} & \numprint{285.22} & \textbf{\numprint{98847}} \\ 
\Id{\detokenize{vermont-AM3}} & \numprint{119.63} & \numprint{63234} & \numprint{88.35} & \numprint{63238} & \numprint{0.95} & \textbf{\numprint{63302}} & \numprint{95.81} & \numprint{55584} & \numprint{443.88} & \numprint{55577} \\ 
\rowcolor{lightergray} \Id{\detokenize{virginia-AM2}} & \numprint{0.89} & \numprint{295794} & \numprint{1.32} & \numprint{295668} & \numprint{0.12} & \textbf{\numprint{295867}} & \numprint{0.93} & \textbf{\numprint{295867}} & \numprint{0.77} & \textbf{\numprint{295867}} \\ 
\Id{\detokenize{virginia-AM3}} & \numprint{289.23} & \numprint{307867} & \numprint{883.75} & \numprint{307845} & \numprint{3.75} & \textbf{\numprint{308305}} & \numprint{109.20} & \numprint{306985} & \numprint{786.05} & \numprint{233572} \\ 
\rowcolor{lightergray} \Id{\detokenize{washington-AM2}} & \numprint{2.00} & \textbf{\numprint{305619}} & \numprint{15.60} & \textbf{\numprint{305619}} & \numprint{0.62} & \textbf{\numprint{305619}} & \numprint{2.44} & \textbf{\numprint{305619}} & \numprint{2.20} & \textbf{\numprint{305619}} \\ 
\Id{\detokenize{washington-AM3}} & \numprint{79.77} & \numprint{313808} & \numprint{401.59} & \numprint{313827} & \numprint{13.88} & \textbf{\numprint{314288}} & \numprint{248.77} & \numprint{271747} & \numprint{532.25} & \numprint{271747} \\ 
\Id{\detokenize{west-virginia-AM3}} & \numprint{1.10} & \textbf{\numprint{47927}} & \numprint{0.87} & \textbf{\numprint{47927}} & \numprint{0.08} & \textbf{\numprint{47927}} & \numprint{14.38} & \textbf{\numprint{47927}} & \numprint{854.73} & \textbf{\numprint{47927}} \\ 
\end{tabular} 

\caption{Best solution found by each algorithm and time (in seconds) required to compute it. The global best solution is highlighted in \textbf{bold}. Rows are highlighted in gray if B~\&~R is able to find an exact solution.
}
\label{tab:osm_local_rt}
\end{table*}

\begin{table*}[ht]
\scriptsize
\centering
\setlength{\tabcolsep}{0.5ex}
\begin{tabular}{l|r r|r r|r r|r r|r r} 
& \multicolumn{2}{c|}{DynWVC1} & \multicolumn{2}{c|}{DynWVC2} & \multicolumn{2}{c|}{HILS} & \multicolumn{2}{c|}{\BRD} & \multicolumn{2}{c}{\BRF} \\ 
\hline 
Graph & $t_\text{max}$ & $w_\text{max}$ & $t_\text{max}$ & $w_\text{max}$ & $t_\text{max}$ & $w_\text{max}$ & $t_\text{max}$ & $w_\text{max}$ & $t_\text{max}$ & $w_\text{max}$ \\ 
\hline 
\Id{\detokenize{as-skitter}} & \numprint{997.39} & \numprint{123412428} & \numprint{576.93} & \numprint{123105765} & \numprint{998.75} & \numprint{122539706} & \numprint{641.38} & \numprint{123172824} & \numprint{746.93} & \textbf{\numprint{123904741}} \\ 
\rowcolor{lightergray} \Id{\detokenize{ca-AstroPh}} & \numprint{207.99} & \numprint{796467} & \numprint{108.35} & \numprint{796535} & \numprint{46.76} & \textbf{\numprint{796556}} & \numprint{0.03} & \textbf{\numprint{796556}} & \numprint{0.03} & \textbf{\numprint{796556}} \\ 
\rowcolor{lightergray} \Id{\detokenize{ca-CondMat}} & \numprint{71.54} & \numprint{1143431} & \numprint{222.30} & \numprint{1143471} & \numprint{45.07} & \textbf{\numprint{1143480}} & \numprint{0.02} & \textbf{\numprint{1143480}} & \numprint{0.02} & \textbf{\numprint{1143480}} \\ 
\rowcolor{lightergray} \Id{\detokenize{ca-GrQc}} & \numprint{1.75} & \textbf{\numprint{289481}} & \numprint{0.82} & \textbf{\numprint{289481}} & \numprint{0.60} & \textbf{\numprint{289481}} & \numprint{0.00} & \textbf{\numprint{289481}} & \numprint{0.00} & \textbf{\numprint{289481}} \\ 
\rowcolor{lightergray} \Id{\detokenize{ca-HepPh}} & \numprint{26.36} & \numprint{579624} & \numprint{17.31} & \numprint{579662} & \numprint{11.44} & \textbf{\numprint{579675}} & \numprint{0.02} & \textbf{\numprint{579675}} & \numprint{0.02} & \textbf{\numprint{579675}} \\ 
\rowcolor{lightergray} \Id{\detokenize{ca-HepTh}} & \numprint{9.87} & \numprint{560630} & \numprint{12.64} & \numprint{560642} & \numprint{94.19} & \textbf{\numprint{560662}} & \numprint{0.01} & \textbf{\numprint{560662}} & \numprint{0.01} & \textbf{\numprint{560662}} \\ 
\rowcolor{lightergray} \Id{\detokenize{email-Enron}} & \numprint{295.02} & \numprint{2457460} & \numprint{910.50} & \numprint{2457505} & \numprint{79.40} & \textbf{\numprint{2457547}} & \numprint{0.04} & \textbf{\numprint{2457547}} & \numprint{0.03} & \textbf{\numprint{2457547}} \\ 
\rowcolor{lightergray} \Id{\detokenize{email-EuAll}} & \numprint{180.92} & \textbf{\numprint{25330331}} & \numprint{179.26} & \textbf{\numprint{25330331}} & \numprint{501.09} & \textbf{\numprint{25330331}} & \numprint{0.13} & \textbf{\numprint{25330331}} & \numprint{0.19} & \textbf{\numprint{25330331}} \\ 
\rowcolor{lightergray} \Id{\detokenize{p2p-Gnutella04}} & \numprint{2.46} & \numprint{667496} & \numprint{866.88} & \numprint{667503} & \numprint{2.64} & \textbf{\numprint{667539}} & \numprint{0.01} & \textbf{\numprint{667539}} & \numprint{0.01} & \textbf{\numprint{667539}} \\ 
\rowcolor{lightergray} \Id{\detokenize{p2p-Gnutella05}} & \numprint{24.23} & \textbf{\numprint{556559}} & \numprint{3.54} & \textbf{\numprint{556559}} & \numprint{0.60} & \textbf{\numprint{556559}} & \numprint{0.01} & \textbf{\numprint{556559}} & \numprint{0.01} & \textbf{\numprint{556559}} \\ 
\rowcolor{lightergray} \Id{\detokenize{p2p-Gnutella06}} & \numprint{532.67} & \numprint{547585} & \numprint{1.38} & \numprint{547586} & \numprint{1.47} & \textbf{\numprint{547591}} & \numprint{0.01} & \textbf{\numprint{547591}} & \numprint{0.01} & \textbf{\numprint{547591}} \\ 
\rowcolor{lightergray} \Id{\detokenize{p2p-Gnutella08}} & \numprint{0.21} & \textbf{\numprint{435893}} & \numprint{0.19} & \textbf{\numprint{435893}} & \numprint{0.25} & \textbf{\numprint{435893}} & \numprint{0.00} & \textbf{\numprint{435893}} & \numprint{0.01} & \textbf{\numprint{435893}} \\ 
\rowcolor{lightergray} \Id{\detokenize{p2p-Gnutella09}} & \numprint{0.23} & \textbf{\numprint{568472}} & \numprint{0.22} & \textbf{\numprint{568472}} & \numprint{0.15} & \textbf{\numprint{568472}} & \numprint{0.01} & \textbf{\numprint{568472}} & \numprint{0.01} & \textbf{\numprint{568472}} \\ 
\rowcolor{lightergray} \Id{\detokenize{p2p-Gnutella24}} & \numprint{10.83} & \numprint{1970325} & \numprint{9.81} & \numprint{1970325} & \numprint{4.06} & \textbf{\numprint{1970329}} & \numprint{0.02} & \textbf{\numprint{1970329}} & \numprint{0.02} & \textbf{\numprint{1970329}} \\ 
\rowcolor{lightergray} \Id{\detokenize{p2p-Gnutella25}} & \numprint{2.22} & \textbf{\numprint{1697310}} & \numprint{6.33} & \textbf{\numprint{1697310}} & \numprint{1.64} & \textbf{\numprint{1697310}} & \numprint{0.01} & \textbf{\numprint{1697310}} & \numprint{0.02} & \textbf{\numprint{1697310}} \\ 
\rowcolor{lightergray} \Id{\detokenize{p2p-Gnutella30}} & \numprint{10.06} & \numprint{2785926} & \numprint{22.66} & \numprint{2785922} & \numprint{7.36} & \textbf{\numprint{2785957}} & \numprint{0.02} & \textbf{\numprint{2785957}} & \numprint{0.03} & \textbf{\numprint{2785957}} \\ 
\rowcolor{lightergray} \Id{\detokenize{p2p-Gnutella31}} & \numprint{169.03} & \numprint{4750622} & \numprint{43.15} & \numprint{4750632} & \numprint{34.33} & \textbf{\numprint{4750671}} & \numprint{0.13} & \textbf{\numprint{4750671}} & \numprint{0.04} & \textbf{\numprint{4750671}} \\ 
\rowcolor{lightergray} \Id{\detokenize{roadNet-CA}} & \numprint{1001.61} & \numprint{109028140} & \numprint{1000.88} & \numprint{109023976} & \numprint{3312.19} & \numprint{108167310} & \numprint{931.36} & \numprint{106500027} & \numprint{774.56} & \textbf{\numprint{111408830}} \\ 
\rowcolor{lightergray} \Id{\detokenize{roadNet-PA}} & \numprint{720.57} & \numprint{60940033} & \numprint{787.59} & \numprint{60940033} & \numprint{998.56} & \numprint{59915775} & \numprint{988.62} & \numprint{58927755} & \numprint{32.06} & \textbf{\numprint{61686106}} \\ 
\rowcolor{lightergray} \Id{\detokenize{roadNet-TX}} & \numprint{1001.45} & \numprint{77498612} & \numprint{1000.78} & \numprint{77525099} & \numprint{1697.13} & \numprint{76366577} & \numprint{870.62} & \numprint{75843903} & \numprint{33.49} & \textbf{\numprint{78606965}} \\ 
\rowcolor{lightergray} \Id{\detokenize{soc-Epinions1}} & \numprint{617.40} & \numprint{5668054} & \numprint{625.89} & \numprint{5668180} & \numprint{694.51} & \numprint{5668382} & \numprint{0.07} & \textbf{\numprint{5668401}} & \numprint{0.11} & \textbf{\numprint{5668401}} \\ 
\Id{\detokenize{soc-LiveJournal1}} & \numprint{1001.31} & \numprint{277850684} & \numprint{1001.23} & \numprint{277824322} & \numprint{12437.50} & \numprint{280559036} & \numprint{86.66} & \numprint{283869420} & \numprint{270.96} & \textbf{\numprint{283948671}} \\ 
\rowcolor{lightergray} \Id{\detokenize{soc-Slashdot0811}} & \numprint{809.97} & \numprint{5650118} & \numprint{477.14} & \numprint{5650303} & \numprint{767.51} & \numprint{5650644} & \numprint{0.10} & \textbf{\numprint{5650791}} & \numprint{0.18} & \textbf{\numprint{5650791}} \\ 
\rowcolor{lightergray} \Id{\detokenize{soc-Slashdot0902}} & \numprint{783.10} & \numprint{5953052} & \numprint{272.11} & \numprint{5953235} & \numprint{786.70} & \numprint{5953436} & \numprint{0.13} & \textbf{\numprint{5953582}} & \numprint{0.21} & \textbf{\numprint{5953582}} \\ 
\Id{\detokenize{soc-pokec-relationships}} & \numprint{999.99} & \numprint{82522272} & \numprint{1001.42} & \textbf{\numprint{82640035}} & \numprint{2482.18} & \numprint{82381583} & \numprint{287.40} & \numprint{82595492} & \numprint{1404.57} & \numprint{75717984} \\ 
\Id{\detokenize{web-BerkStan}} & \numprint{347.17} & \numprint{43595139} & \numprint{372.33} & \numprint{43593142} & \numprint{994.73} & \numprint{43319988} & \numprint{22.58} & \numprint{43138612} & \numprint{831.75} & \textbf{\numprint{43766431}} \\ 
\rowcolor{lightergray} \Id{\detokenize{web-Google}} & \numprint{759.75} & \numprint{56193138} & \numprint{683.63} & \numprint{56190870} & \numprint{994.58} & \numprint{55954155} & \numprint{2.08} & \textbf{\numprint{56313384}} & \numprint{3.16} & \textbf{\numprint{56313384}} \\ 
\Id{\detokenize{web-NotreDame}} & \numprint{963.44} & \textbf{\numprint{25975765}} & \numprint{875.22} & \numprint{25968209} & \numprint{998.79} & \numprint{25970368} & \numprint{354.79} & \numprint{25947936} & \numprint{28.11} & \numprint{25957800} \\ 
\Id{\detokenize{web-Stanford}} & \numprint{999.97} & \numprint{17731195} & \numprint{997.98} & \numprint{17735700} & \numprint{999.91} & \numprint{17679156} & \numprint{47.62} & \numprint{17634819} & \numprint{4.69} & \textbf{\numprint{17799469}} \\ 
\rowcolor{lightergray} \Id{\detokenize{wiki-Talk}} & \numprint{961.05} & \numprint{235874406} & \numprint{991.31} & \numprint{235874419} & \numprint{996.02} & \numprint{235852509} & \numprint{3.85} & \textbf{\numprint{235875181}} & \numprint{3.36} & \textbf{\numprint{235875181}} \\ 
\rowcolor{lightergray} \Id{\detokenize{wiki-Vote}} & \numprint{0.74} & \textbf{\numprint{500436}} & \numprint{0.75} & \textbf{\numprint{500436}} & \numprint{23.96} & \textbf{\numprint{500436}} & \numprint{0.05} & \textbf{\numprint{500436}} & \numprint{0.06} & \textbf{\numprint{500436}} \\ 
\end{tabular} 

\caption{Best solution found by each algorithm and time (in seconds) required to compute it. The global best solution is highlighted in \textbf{bold}. Rows are highlighted in gray if B~\&~R is able to find an exact solution.
}
\label{tab:snap_local_rt}
\end{table*}

\begin{table*}[ht]
\scriptsize
\centering
\setlength{\tabcolsep}{0.5ex}
\begin{tabular}{l|r r|r r|r r|r r|r r} 
& \multicolumn{2}{c|}{Red+DynWVC1} & \multicolumn{2}{c|}{Red+DynWVC2} & \multicolumn{2}{c|}{Red+HILS} & \multicolumn{2}{c|}{\BRD} & \multicolumn{2}{c}{\BRF} \\ 
\hline 
Graph & $t_\text{max}$ & $w_\text{max}$ & $t_\text{max}$ & $w_\text{max}$ & $t_\text{max}$ & $w_\text{max}$ & $t_\text{max}$ & $w_\text{max}$ & $t_\text{max}$ & $w_\text{max}$ \\ 
\hline 
\rowcolor{lightergray} \Id{\detokenize{alabama-AM2}} & \numprint{0.11} & \textbf{\numprint{174309}} & \numprint{0.11} & \textbf{\numprint{174309}} & \numprint{0.10} & \textbf{\numprint{174309}} & \numprint{0.40} & \textbf{\numprint{174309}} & \numprint{0.79} & \textbf{\numprint{174309}} \\ 
\Id{\detokenize{alabama-AM3}} & \numprint{370.80} & \numprint{185727} & \numprint{295.20} & \numprint{185729} & \numprint{4.05} & \textbf{\numprint{185744}} & \numprint{15.79} & \numprint{185707} & \numprint{80.78} & \numprint{185707} \\ 
\Id{\detokenize{district-of-columbia-AM1}} & \numprint{0.92} & \textbf{\numprint{196475}} & \numprint{0.92} & \textbf{\numprint{196475}} & \numprint{0.37} & \textbf{\numprint{196475}} & \numprint{1.97} & \textbf{\numprint{196475}} & \numprint{4.13} & \textbf{\numprint{196475}} \\ 
\Id{\detokenize{district-of-columbia-AM2}} & \numprint{334.12} & \numprint{209125} & \numprint{982.91} & \numprint{209056} & \numprint{220.82} & \textbf{\numprint{209132}} & \numprint{20.03} & \numprint{147450} & \numprint{233.70} & \numprint{147450} \\ 
\Id{\detokenize{district-of-columbia-AM3}} & \numprint{879.25} & \numprint{225535} & \numprint{789.47} & \numprint{225031} & \numprint{320.06} & \textbf{\numprint{227534}} & \numprint{553.84} & \numprint{92784} & \numprint{918.07} & \numprint{92714} \\ 
\rowcolor{lightergray} \Id{\detokenize{florida-AM2}} & \numprint{0.03} & \textbf{\numprint{230595}} & \numprint{0.03} & \textbf{\numprint{230595}} & \numprint{0.03} & \textbf{\numprint{230595}} & \numprint{0.03} & \textbf{\numprint{230595}} & \numprint{0.02} & \textbf{\numprint{230595}} \\ 
\rowcolor{lightergray} \Id{\detokenize{florida-AM3}} & \numprint{8.66} & \numprint{237331} & \numprint{8.57} & \numprint{237331} & \numprint{8.01} & \textbf{\numprint{237333}} & \numprint{20.52} & \textbf{\numprint{237333}} & \numprint{324.38} & \numprint{226767} \\ 
\Id{\detokenize{georgia-AM3}} & \numprint{2.64} & \textbf{\numprint{222652}} & \numprint{2.62} & \textbf{\numprint{222652}} & \numprint{2.43} & \textbf{\numprint{222652}} & \numprint{4.88} & \numprint{214918} & \numprint{14.35} & \numprint{214918} \\ 
\Id{\detokenize{greenland-AM3}} & \numprint{712.63} & \numprint{14007} & \numprint{462.23} & \numprint{14006} & \numprint{10.34} & \textbf{\numprint{14011}} & \numprint{14.52} & \numprint{13152} & \numprint{47.25} & \numprint{13069} \\ 
\rowcolor{lightergray} \Id{\detokenize{hawaii-AM2}} & \numprint{0.96} & \textbf{\numprint{125284}} & \numprint{0.96} & \textbf{\numprint{125284}} & \numprint{0.93} & \textbf{\numprint{125284}} & \numprint{3.59} & \textbf{\numprint{125284}} & \numprint{10.89} & \textbf{\numprint{125284}} \\ 
\Id{\detokenize{hawaii-AM3}} & \numprint{405.34} & \numprint{140714} & \numprint{957.61} & \numprint{140709} & \numprint{329.20} & \textbf{\numprint{141011}} & \numprint{288.58} & \numprint{106251} & \numprint{1177.95} & \numprint{129812} \\ 
\Id{\detokenize{idaho-AM3}} & \numprint{40.38} & \textbf{\numprint{77145}} & \numprint{20.79} & \textbf{\numprint{77145}} & \numprint{203.76} & \textbf{\numprint{77145}} & \numprint{866.90} & \numprint{77010} & \numprint{61.26} & \numprint{76831} \\ 
\Id{\detokenize{kansas-AM3}} & \numprint{13.59} & \textbf{\numprint{87976}} & \numprint{18.43} & \textbf{\numprint{87976}} & \numprint{2.06} & \textbf{\numprint{87976}} & \numprint{11.35} & \numprint{87925} & \numprint{18.99} & \numprint{87925} \\ 
\rowcolor{lightergray} \Id{\detokenize{kentucky-AM2}} & \numprint{1.13} & \textbf{\numprint{97397}} & \numprint{1.13} & \textbf{\numprint{97397}} & \numprint{1.07} & \textbf{\numprint{97397}} & \numprint{11.35} & \textbf{\numprint{97397}} & \numprint{42.05} & \textbf{\numprint{97397}} \\ 
\Id{\detokenize{kentucky-AM3}} & \numprint{766.39} & \numprint{100479} & \numprint{759.20} & \numprint{100480} & \numprint{973.22} & \textbf{\numprint{100486}} & \numprint{172.30} & \numprint{91864} & \numprint{3346.94} & \numprint{96634} \\ 
\rowcolor{lightergray} \Id{\detokenize{louisiana-AM3}} & \numprint{1.35} & \textbf{\numprint{60024}} & \numprint{1.35} & \textbf{\numprint{60024}} & \numprint{1.33} & \textbf{\numprint{60024}} & \numprint{3.38} & \textbf{\numprint{60024}} & \numprint{20.17} & \textbf{\numprint{60024}} \\ 
\rowcolor{lightergray} \Id{\detokenize{maryland-AM3}} & \numprint{2.07} & \textbf{\numprint{45496}} & \numprint{2.07} & \textbf{\numprint{45496}} & \numprint{2.07} & \textbf{\numprint{45496}} & \numprint{3.34} & \textbf{\numprint{45496}} & \numprint{11.08} & \textbf{\numprint{45496}} \\ 
\rowcolor{lightergray} \Id{\detokenize{massachusetts-AM2}} & \numprint{0.04} & \textbf{\numprint{140095}} & \numprint{0.04} & \textbf{\numprint{140095}} & \numprint{0.04} & \textbf{\numprint{140095}} & \numprint{0.46} & \textbf{\numprint{140095}} & \numprint{0.48} & \textbf{\numprint{140095}} \\ 
\Id{\detokenize{massachusetts-AM3}} & \numprint{10.68} & \textbf{\numprint{145866}} & \numprint{8.38} & \textbf{\numprint{145866}} & \numprint{2.92} & \textbf{\numprint{145866}} & \numprint{12.87} & \numprint{145617} & \numprint{23.97} & \numprint{145631} \\ 
\rowcolor{lightergray} \Id{\detokenize{mexico-AM3}} & \numprint{5.39} & \textbf{\numprint{97663}} & \numprint{5.34} & \textbf{\numprint{97663}} & \numprint{5.28} & \textbf{\numprint{97663}} & \numprint{14.25} & \textbf{\numprint{97663}} & \numprint{289.14} & \textbf{\numprint{97663}} \\ 
\rowcolor{lightergray} \Id{\detokenize{new-hampshire-AM3}} & \numprint{1.51} & \textbf{\numprint{116060}} & \numprint{1.51} & \textbf{\numprint{116060}} & \numprint{1.50} & \textbf{\numprint{116060}} & \numprint{3.25} & \textbf{\numprint{116060}} & \numprint{8.75} & \textbf{\numprint{116060}} \\ 
\Id{\detokenize{north-carolina-AM3}} & \numprint{1.76} & \textbf{\numprint{49720}} & \numprint{0.79} & \textbf{\numprint{49720}} & \numprint{0.48} & \textbf{\numprint{49720}} & \numprint{10.45} & \numprint{49562} & \numprint{11.55} & \numprint{49562} \\ 
\rowcolor{lightergray} \Id{\detokenize{oregon-AM2}} & \numprint{0.04} & \textbf{\numprint{165047}} & \numprint{0.04} & \textbf{\numprint{165047}} & \numprint{0.04} & \textbf{\numprint{165047}} & \numprint{0.04} & \textbf{\numprint{165047}} & \numprint{0.09} & \textbf{\numprint{165047}} \\ 
\Id{\detokenize{oregon-AM3}} & \numprint{135.72} & \numprint{175073} & \numprint{167.56} & \numprint{175075} & \numprint{5.18} & \textbf{\numprint{175078}} & \numprint{351.99} & \numprint{174334} & \numprint{474.15} & \numprint{164941} \\ 
\rowcolor{lightergray} \Id{\detokenize{pennsylvania-AM3}} & \numprint{4.35} & \textbf{\numprint{143870}} & \numprint{4.34} & \textbf{\numprint{143870}} & \numprint{4.33} & \textbf{\numprint{143870}} & \numprint{9.98} & \textbf{\numprint{143870}} & \numprint{38.76} & \textbf{\numprint{143870}} \\ 
\Id{\detokenize{rhode-island-AM2}} & \numprint{1.03} & \textbf{\numprint{184596}} & \numprint{2.40} & \textbf{\numprint{184596}} & \numprint{0.43} & \textbf{\numprint{184596}} & \numprint{10.70} & \numprint{184543} & \numprint{16.79} & \numprint{184543} \\ 
\Id{\detokenize{rhode-island-AM3}} & \numprint{993.86} & \numprint{201667} & \numprint{255.71} & \numprint{201668} & \numprint{605.61} & \textbf{\numprint{201734}} & \numprint{399.33} & \numprint{162639} & \numprint{931.05} & \numprint{163080} \\ 
\rowcolor{lightergray} \Id{\detokenize{utah-AM3}} & \numprint{168.07} & \textbf{\numprint{98847}} & \numprint{2.36} & \textbf{\numprint{98847}} & \numprint{2.10} & \textbf{\numprint{98847}} & \numprint{64.04} & \textbf{\numprint{98847}} & \numprint{285.22} & \textbf{\numprint{98847}} \\ 
\Id{\detokenize{vermont-AM3}} & \numprint{62.85} & \numprint{63280} & \numprint{690.58} & \numprint{63256} & \numprint{2.95} & \textbf{\numprint{63312}} & \numprint{95.81} & \numprint{55584} & \numprint{443.88} & \numprint{55577} \\ 
\rowcolor{lightergray} \Id{\detokenize{virginia-AM2}} & \numprint{0.25} & \textbf{\numprint{295867}} & \numprint{0.25} & \textbf{\numprint{295867}} & \numprint{0.23} & \textbf{\numprint{295867}} & \numprint{0.93} & \textbf{\numprint{295867}} & \numprint{0.77} & \textbf{\numprint{295867}} \\ 
\Id{\detokenize{virginia-AM3}} & \numprint{708.66} & \numprint{308052} & \numprint{790.21} & \numprint{308090} & \numprint{19.34} & \textbf{\numprint{308305}} & \numprint{109.20} & \numprint{306985} & \numprint{786.05} & \numprint{233572} \\ 
\rowcolor{lightergray} \Id{\detokenize{washington-AM2}} & \numprint{0.24} & \textbf{\numprint{305619}} & \numprint{0.24} & \textbf{\numprint{305619}} & \numprint{0.23} & \textbf{\numprint{305619}} & \numprint{2.44} & \textbf{\numprint{305619}} & \numprint{2.20} & \textbf{\numprint{305619}} \\ 
\Id{\detokenize{washington-AM3}} & \numprint{59.08} & \numprint{314097} & \numprint{505.58} & \numprint{314079} & \numprint{863.47} & \textbf{\numprint{314288}} & \numprint{248.77} & \numprint{271747} & \numprint{532.25} & \numprint{271747} \\ 
\Id{\detokenize{west-virginia-AM3}} & \numprint{3.06} & \textbf{\numprint{47927}} & \numprint{3.77} & \textbf{\numprint{47927}} & \numprint{2.54} & \textbf{\numprint{47927}} & \numprint{14.38} & \textbf{\numprint{47927}} & \numprint{854.73} & \textbf{\numprint{47927}} \\ 
\end{tabular} 

\caption{Best solution found by each algorithm and time (in seconds) required to compute it. The global best solution is highlighted in \textbf{bold}. Rows are highlighted in gray if B~\&~R is able to find an exact solution.
}
\label{tab:osm_reduce_rt}
\end{table*}

\begin{table*}[ht]
\scriptsize
\centering
\setlength{\tabcolsep}{0.5ex}
\begin{tabular}{l|r r|r r|r r|r r|r r} 
& \multicolumn{2}{c|}{Red+DynWVC1} & \multicolumn{2}{c|}{Red+DynWVC2} & \multicolumn{2}{c|}{Red+HILS} & \multicolumn{2}{c|}{\BRD} & \multicolumn{2}{c}{\BRF} \\ 
\hline 
Graph & $t_\text{max}$ & $w_\text{max}$ & $t_\text{max}$ & $w_\text{max}$ & $t_\text{max}$ & $w_\text{max}$ & $t_\text{max}$ & $w_\text{max}$ & $t_\text{max}$ & $w_\text{max}$ \\ 
\hline 
\Id{\detokenize{as-skitter}} & \numprint{64.52} & \numprint{123995654} & \numprint{85.60} & \numprint{123995808} & \numprint{845.70} & \textbf{\numprint{123996322}} & \numprint{641.38} & \numprint{123172824} & \numprint{746.93} & \numprint{123904741} \\ 
\rowcolor{lightergray} \Id{\detokenize{ca-AstroPh}} & \numprint{0.02} & \textbf{\numprint{796556}} & \numprint{0.02} & \textbf{\numprint{796556}} & \numprint{0.02} & \textbf{\numprint{796556}} & \numprint{0.03} & \textbf{\numprint{796556}} & \numprint{0.03} & \textbf{\numprint{796556}} \\ 
\rowcolor{lightergray} \Id{\detokenize{ca-CondMat}} & \numprint{0.01} & \textbf{\numprint{1143480}} & \numprint{0.01} & \textbf{\numprint{1143480}} & \numprint{0.01} & \textbf{\numprint{1143480}} & \numprint{0.02} & \textbf{\numprint{1143480}} & \numprint{0.02} & \textbf{\numprint{1143480}} \\ 
\rowcolor{lightergray} \Id{\detokenize{ca-GrQc}} & \numprint{0.00} & \textbf{\numprint{289481}} & \numprint{0.00} & \textbf{\numprint{289481}} & \numprint{0.00} & \textbf{\numprint{289481}} & \numprint{0.00} & \textbf{\numprint{289481}} & \numprint{0.00} & \textbf{\numprint{289481}} \\ 
\rowcolor{lightergray} \Id{\detokenize{ca-HepPh}} & \numprint{0.02} & \textbf{\numprint{579675}} & \numprint{0.02} & \textbf{\numprint{579675}} & \numprint{0.02} & \textbf{\numprint{579675}} & \numprint{0.02} & \textbf{\numprint{579675}} & \numprint{0.02} & \textbf{\numprint{579675}} \\ 
\rowcolor{lightergray} \Id{\detokenize{ca-HepTh}} & \numprint{0.00} & \textbf{\numprint{560662}} & \numprint{0.00} & \textbf{\numprint{560662}} & \numprint{0.00} & \textbf{\numprint{560662}} & \numprint{0.01} & \textbf{\numprint{560662}} & \numprint{0.01} & \textbf{\numprint{560662}} \\ 
\rowcolor{lightergray} \Id{\detokenize{email-Enron}} & \numprint{0.03} & \textbf{\numprint{2457547}} & \numprint{0.03} & \textbf{\numprint{2457547}} & \numprint{0.03} & \textbf{\numprint{2457547}} & \numprint{0.04} & \textbf{\numprint{2457547}} & \numprint{0.03} & \textbf{\numprint{2457547}} \\ 
\rowcolor{lightergray} \Id{\detokenize{email-EuAll}} & \numprint{0.12} & \textbf{\numprint{25330331}} & \numprint{0.12} & \textbf{\numprint{25330331}} & \numprint{0.12} & \textbf{\numprint{25330331}} & \numprint{0.13} & \textbf{\numprint{25330331}} & \numprint{0.19} & \textbf{\numprint{25330331}} \\ 
\rowcolor{lightergray} \Id{\detokenize{p2p-Gnutella04}} & \numprint{0.01} & \textbf{\numprint{667539}} & \numprint{0.01} & \textbf{\numprint{667539}} & \numprint{0.01} & \textbf{\numprint{667539}} & \numprint{0.01} & \textbf{\numprint{667539}} & \numprint{0.01} & \textbf{\numprint{667539}} \\ 
\rowcolor{lightergray} \Id{\detokenize{p2p-Gnutella05}} & \numprint{0.01} & \textbf{\numprint{556559}} & \numprint{0.01} & \textbf{\numprint{556559}} & \numprint{0.01} & \textbf{\numprint{556559}} & \numprint{0.01} & \textbf{\numprint{556559}} & \numprint{0.01} & \textbf{\numprint{556559}} \\ 
\rowcolor{lightergray} \Id{\detokenize{p2p-Gnutella06}} & \numprint{0.01} & \textbf{\numprint{547591}} & \numprint{0.01} & \textbf{\numprint{547591}} & \numprint{0.01} & \textbf{\numprint{547591}} & \numprint{0.01} & \textbf{\numprint{547591}} & \numprint{0.01} & \textbf{\numprint{547591}} \\ 
\rowcolor{lightergray} \Id{\detokenize{p2p-Gnutella08}} & \numprint{0.00} & \textbf{\numprint{435893}} & \numprint{0.00} & \textbf{\numprint{435893}} & \numprint{0.00} & \textbf{\numprint{435893}} & \numprint{0.00} & \textbf{\numprint{435893}} & \numprint{0.01} & \textbf{\numprint{435893}} \\ 
\rowcolor{lightergray} \Id{\detokenize{p2p-Gnutella09}} & \numprint{0.01} & \textbf{\numprint{568472}} & \numprint{0.01} & \textbf{\numprint{568472}} & \numprint{0.01} & \textbf{\numprint{568472}} & \numprint{0.01} & \textbf{\numprint{568472}} & \numprint{0.01} & \textbf{\numprint{568472}} \\ 
\rowcolor{lightergray} \Id{\detokenize{p2p-Gnutella24}} & \numprint{0.01} & \textbf{\numprint{1970329}} & \numprint{0.01} & \textbf{\numprint{1970329}} & \numprint{0.01} & \textbf{\numprint{1970329}} & \numprint{0.02} & \textbf{\numprint{1970329}} & \numprint{0.02} & \textbf{\numprint{1970329}} \\ 
\rowcolor{lightergray} \Id{\detokenize{p2p-Gnutella25}} & \numprint{0.01} & \textbf{\numprint{1697310}} & \numprint{0.01} & \textbf{\numprint{1697310}} & \numprint{0.01} & \textbf{\numprint{1697310}} & \numprint{0.01} & \textbf{\numprint{1697310}} & \numprint{0.02} & \textbf{\numprint{1697310}} \\ 
\rowcolor{lightergray} \Id{\detokenize{p2p-Gnutella30}} & \numprint{0.01} & \textbf{\numprint{2785957}} & \numprint{0.01} & \textbf{\numprint{2785957}} & \numprint{0.01} & \textbf{\numprint{2785957}} & \numprint{0.02} & \textbf{\numprint{2785957}} & \numprint{0.03} & \textbf{\numprint{2785957}} \\ 
\rowcolor{lightergray} \Id{\detokenize{p2p-Gnutella31}} & \numprint{0.02} & \textbf{\numprint{4750671}} & \numprint{0.02} & \textbf{\numprint{4750671}} & \numprint{0.02} & \textbf{\numprint{4750671}} & \numprint{0.13} & \textbf{\numprint{4750671}} & \numprint{0.04} & \textbf{\numprint{4750671}} \\ 
\rowcolor{lightergray} \Id{\detokenize{roadNet-CA}} & \numprint{918.32} & \numprint{111398659} & \numprint{866.70} & \numprint{111398243} & \numprint{994.57} & \numprint{111402080} & \numprint{931.36} & \numprint{106500027} & \numprint{774.56} & \textbf{\numprint{111408830}} \\ 
\rowcolor{lightergray} \Id{\detokenize{roadNet-PA}} & \numprint{733.57} & \numprint{61680822} & \numprint{639.56} & \numprint{61680822} & \numprint{947.93} & \numprint{61682180} & \numprint{988.62} & \numprint{58927755} & \numprint{32.06} & \textbf{\numprint{61686106}} \\ 
\rowcolor{lightergray} \Id{\detokenize{roadNet-TX}} & \numprint{952.53} & \numprint{78601859} & \numprint{771.05} & \numprint{78601813} & \numprint{946.32} & \numprint{78602984} & \numprint{870.62} & \numprint{75843903} & \numprint{33.49} & \textbf{\numprint{78606965}} \\ 
\rowcolor{lightergray} \Id{\detokenize{soc-Epinions1}} & \numprint{0.08} & \textbf{\numprint{5668401}} & \numprint{0.08} & \textbf{\numprint{5668401}} & \numprint{0.08} & \textbf{\numprint{5668401}} & \numprint{0.07} & \textbf{\numprint{5668401}} & \numprint{0.11} & \textbf{\numprint{5668401}} \\ 
\Id{\detokenize{soc-LiveJournal1}} & \numprint{916.65} & \numprint{283973802} & \numprint{996.68} & \numprint{283973997} & \numprint{761.51} & \textbf{\numprint{283975036}} & \numprint{86.66} & \numprint{283869420} & \numprint{270.96} & \numprint{283948671} \\ 
\rowcolor{lightergray} \Id{\detokenize{soc-Slashdot0811}} & \numprint{0.14} & \textbf{\numprint{5650791}} & \numprint{0.14} & \textbf{\numprint{5650791}} & \numprint{0.14} & \textbf{\numprint{5650791}} & \numprint{0.10} & \textbf{\numprint{5650791}} & \numprint{0.18} & \textbf{\numprint{5650791}} \\ 
\rowcolor{lightergray} \Id{\detokenize{soc-Slashdot0902}} & \numprint{0.17} & \textbf{\numprint{5953582}} & \numprint{0.17} & \textbf{\numprint{5953582}} & \numprint{0.17} & \textbf{\numprint{5953582}} & \numprint{0.13} & \textbf{\numprint{5953582}} & \numprint{0.21} & \textbf{\numprint{5953582}} \\ 
\Id{\detokenize{soc-pokec-relationships}} & \numprint{1400.47} & \numprint{43734005} & \numprint{1400.47} & \numprint{43734005} & \numprint{2400.00} & \textbf{\numprint{82845330}} & \numprint{287.40} & \numprint{82595492} & \numprint{1404.57} & \numprint{75717984} \\ 
\Id{\detokenize{web-BerkStan}} & \numprint{373.58} & \numprint{43877439} & \numprint{612.64} & \numprint{43877349} & \numprint{859.76} & \textbf{\numprint{43877507}} & \numprint{22.58} & \numprint{43138612} & \numprint{831.75} & \numprint{43766431} \\ 
\rowcolor{lightergray} \Id{\detokenize{web-Google}} & \numprint{3.20} & \numprint{56313343} & \numprint{3.30} & \numprint{56313349} & \numprint{3.01} & \textbf{\numprint{56313384}} & \numprint{2.08} & \textbf{\numprint{56313384}} & \numprint{3.16} & \textbf{\numprint{56313384}} \\ 
\Id{\detokenize{web-NotreDame}} & \numprint{147.60} & \numprint{25995575} & \numprint{850.00} & \numprint{25995615} & \numprint{173.50} & \textbf{\numprint{25995648}} & \numprint{354.79} & \numprint{25947936} & \numprint{28.11} & \numprint{25957800} \\ 
\Id{\detokenize{web-Stanford}} & \numprint{5.08} & \numprint{17799379} & \numprint{5.19} & \numprint{17799405} & \numprint{131.24} & \textbf{\numprint{17799556}} & \numprint{47.62} & \numprint{17634819} & \numprint{4.69} & \numprint{17799469} \\ 
\rowcolor{lightergray} \Id{\detokenize{wiki-Talk}} & \numprint{2.30} & \textbf{\numprint{235875181}} & \numprint{2.30} & \textbf{\numprint{235875181}} & \numprint{2.30} & \textbf{\numprint{235875181}} & \numprint{3.85} & \textbf{\numprint{235875181}} & \numprint{3.36} & \textbf{\numprint{235875181}} \\ 
\rowcolor{lightergray} \Id{\detokenize{wiki-Vote}} & \numprint{0.04} & \textbf{\numprint{500436}} & \numprint{0.04} & \textbf{\numprint{500436}} & \numprint{0.03} & \textbf{\numprint{500436}} & \numprint{0.05} & \textbf{\numprint{500436}} & \numprint{0.06} & \textbf{\numprint{500436}} \\ 
\end{tabular} 

\caption{Best solution found by each algorithm and time (in seconds) required to compute it. The global best solution is highlighted in \textbf{bold}. Rows are highlighted in gray if B~\&~R is able to find an exact solution.
}
\label{tab:snap_reduce_rt}
\end{table*}

\begin{figure*}[t]
\centering
\includegraphics[scale=0.45]{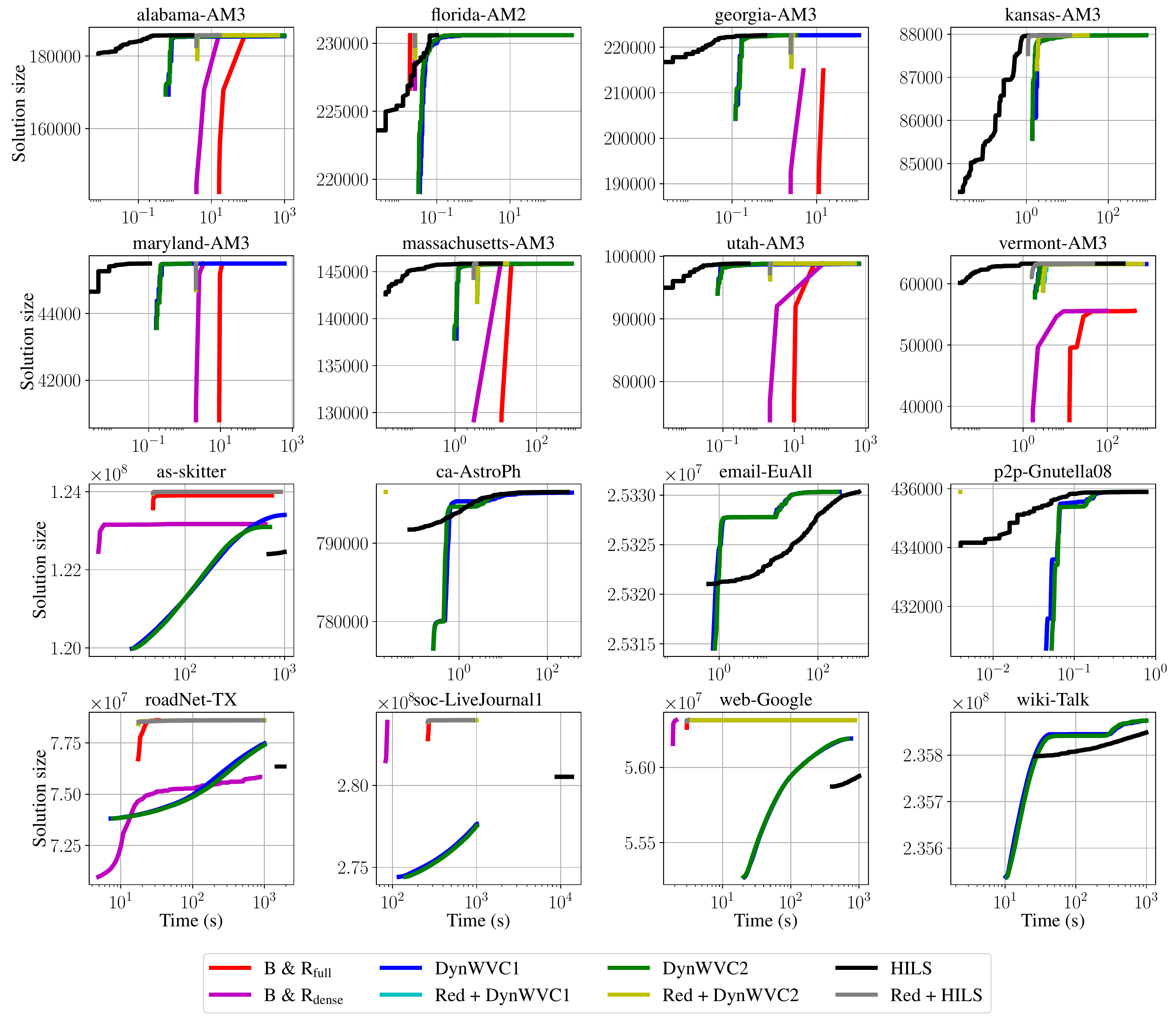}
\caption{Solution quality over time for our sample of eight OSM instances (upper two rows) and eight SNAP instances (lower two rows) as given in Section~\ref{sec:sota}.}
\label{fig:apdxconvergence}
\end{figure*}

\end{appendix}
\end{document}